%% file: ms.tex
\documentclass[12pt,preprint]{aastex}

\usepackage{graphicx, natbib, psfig}
\shortauthors{Glikman, Helfand \& White}
\shorttitle{Near-infrared Quasar Template}

\begin{document}
\title {A Near-Infrared Spectral Template for Quasars}

\author{Eilat Glikman\altaffilmark{1}}
\email{eilatg@astro.columbia.edu}
\author{David J. Helfand\altaffilmark{1}}
\email{djh@astro.columbia.edu}
\and
\author{Richard L. White\altaffilmark{2}}
\email{rlw@stsci.edu}

\altaffiltext{1}{Columbia Univeristy Department of Astronomy, 550 West 120th Street, New York, NY 10027}
\altaffiltext{2}{Space Telescope Science Institute, 3700 San Martin Dr., Baltimore, MD 21218}

\begin{abstract}
We present a near-infrared quasar composite spectrum spanning the wavelength range 0.58 - 3.5 \micron.  The spectrum has been constructed from observations of 27 quasars obtained at the NASA IRTF telescope and satisfying the criteria $K_s < 14.5$ and $M_i < -23$; the redshift range is $0.118 < z < 0.418$.  The signal-to-noise is moderate, reaching a maximum of 150 between 1.6 and 1.9 \micron.  While a power-law fit to the continuum of the composite spectrum requires two breaks, a single power-law slope of $\alpha=-0.92$ plus a 1260 K blackbody provides an excellent description of the spectrum from H$\alpha$ to 3.5 \micron, strongly suggesting the presence of significant quantities of hot dust in this blue-selected quasar sample.  We measure intensities and line widths for ten lines, finding that the Paschen line ratios rule out Case B recombination.  We compute $K$-corrections for the $J$, $H$, $K$, and {\em Spitzer} 3.6 \micron\ bands which will be useful in analyzing observations of quasars up to $z=10$.
\end{abstract}

\section {Introduction}
Spectral templates constructed from large uniform source samples have been invaluable for understanding the class of objects in question.  Composite spectra enhance common features, while unusual characteristics are averaged out \citep[see discussions in ][]{Scott04,Telfer02,VandenBerk01,Brotherton01,Zheng97,Francis91,Boyle90,Cristiani90}.  For quasars in the ultraviolet to optical regimes, comparison of individual spectra with such composites helps to distinguish normal features from those which arise from unusual physical conditions.  The average shape of the rest-frame UV continuum has provided insight into the source of photoionization of the quasar's emission line gas, while the emission lines provide information on temperatures and densities and have helped discriminate among AGN models. 

Composite spectra can also be useful in quasar searches, and in
comparing new subclasses of sources with existing populations.
For example, we \citep{Glikman04} have recently sought to assess the
incompleteness of the blue-selected quasar population by searching
for highly obscured and/or very red quasars among a candidate list
derived by matching the FIRST (radio) and 2MASS (near-IR) surveys.
We found seventeen highly obscured quasars (E(B-V)$>0.5$) with
red spectral energy distributions by obtaining near-IR spectra spanning
the wavelength range 0.9-2.5$\mu$m. Since existing quasar composite spectra
extend only to $\sim 8500$ \AA, they are of little use in interpreting the
line and continuum emission in these obscured objects for $z<1$. This
motivated us to construct a near-infrared composite spectrum for a
sample of normal quasars with blue colors and little or no obscuration
in order to be able to assess how similar our red-selected sample is
to the dominant population. In addition, our IR composite will prove useful
when observing high-redshift objects with the InfraRed Spectrograph (IRS)
onboard Spitzer whose wavelength coverage extends from 5 to 35$\mu$m,
allowing comparisons for objects out to $z\sim 4-5$.

In this work we present the first near-infrared quasar composite spectrum covering the 0.58-3.5\micron\ rest wavelength range.  Our quasars were chosen to be luminous, low-redshift, UVX-selected quasars from the Sloan Digital Sky Survey Quasar Catalog First Data Release (SDSS-DR1) catalog.  We discuss the details of our selection criteria and the observations in \S2.  We describe the method for constructing the composite and present the results in \S3.  In \S4 we analyze the continuum and line features in the spectrum followed by a comparison with previous infrared quasar SEDs from the literature (\S5).  In \S6 we present a combined optical-to-near-infrared composite, utilizing the SDSS optical spectrum available for each source, and use the resultant spectrum to compute $K$-corrections for the 2MASS $J$, $H$, and $K$ bands, as well as for the {\em Spitzer} 3.6 \micron\ IRAC band.  We summarize this work in \S7 and discuss future efforts to extend and improve our knowledge of the intrinsic quasar spectrum.  We use standard cosmological parameters throughout the paper: $H_0=70$ km s$^{-1}$ Mpc$^{-1}$, $\Omega_M=0.30$, $\Omega_\Lambda=0.70$.

\section {Object Selection and Observations}

The quasars selected for inclusion in this composite spectrum were chosen from the SDSS-DR1 quasar catalog \citep{Schneider03} which contains 16,713 objects.  Of these, 2260 had detections in the Two Micron All Sky Survey (2MASS, Strutskie et al. 1997) Point Source Catalog (PSC).  In order to optimize the use of telescope time by obtaining many spectra with good signal-to-noise ratios, we selected objects with $K_s < 14.5$.  We restrict our objects to relatively low redshifts ($z<0.5$) to focus on the quasars' rest-frame near-infrared emission;  higher redshift objects will contain optical lines which have already been explored in optical quasar surveys.  Since we would like to avoid a significant contribution from the quasar's host galaxy in the $K$ band, we require that our objects have absolute $i$-band magnitudes of $M_i < -23$.  With this choice, an $L_{i}^\star$ host will contribute at most $40\%$ of the total $i$-band flux, assuming $M_{i}^\star = -22.03$ \citep{Blanton01}; with our projected slit widths (see below), the galaxy fraction is $\lesssim 8\%$.  There are 87 objects in the DR1 Quasar catalog which meet these criteria. 

We obtained near-infrared spectra for 27 of the quasars obeying the above criteria.  The observations are summarized in Table \ref{observations}.  All spectra but one were obtained at the NASA Infrared Telescope Facility (IRTF), a 3.0 meter telescope optimized for infrared observations on the summit of Mauna Kea, Hawai`i.  We used Spex \citep{Rayner03} in short and long cross-dispersed modes (SXD, 0.8-2.5 \micron\ and LXD, 1.9-4.1 \micron, respectively), obtaining 24 SXD and 13 LXD spectra with $\lambda/(\Delta \lambda) \simeq 1200$ and $1500$, respectively. The data in each mode were reduced using Spextool \citep*{Cushing04} and corrected for telluric absorption and instrument response with the Telluric software package \citep*{Vacca03}; a standard A0V star spectrum was obtained immediately after each object spectrum using calibrators that lay within both a difference in airmass of 0.1 and a separation from the target of $< 10\degr$.  The SXD and LXD spectra were merged using the MERGEXD task within the Spextool package which scales the weaker spectrum to the brighter spectrum using a selected region in the $K$ band where the two modes overlap.  The fully reduced spectra are available in the online version of the article.  The IRTF data were obtained between 2003 September 23 and 2004 September 14 over three observing runs.  Seeing ranged from $\sim$ 0\farcs6 to $\sim$ 1\farcs25 FWHM; we mostly used the 0\farcs5 and 0\farcs8 slits as appropriate (see columns (6) and (9) of Table \ref{observations}).  For all but a few of the SXD spectra, we used the 0\farcs5 slit which projects $1.5-2.8$ kpc on the sky in our redshift range $0.18-0.42$.  This is at least an order of magnitude larger than the size of the typical narrow-line region of a quasar ($\sim 100$ pc) and we therefore are inevitably including in a large amount of host galaxy light.  Assuming a de Vaucouleurs ($r^{1/4}$-law) surface-density profile with a half-light radius of 11.5 kpc \citep{Dunlop03} and depending on the quasar's redshift, this corresponds to $13-20\%$ of the total host galaxy light.  We discuss in greater detail the impact of host galaxy light on our composite in Sections 4.1 and 6.

One spectrum was obtained at the MDM Observatory on Kitt Peak on 2004 November 24 with the TIFKAM infrared camera in spectroscopic mode using a 0\farcs6 slit .  Conditions were photometric and the seeing was 0\farcs9 FWHM.  This spectrum spans 0.95-2.45\micron\ at a lower resolving power of 600.  The spectrum was extracted and reduced using standard spectral reduction tasks in IRAF.  The telluric absorption and instrument response were corrected using a general version of the Telluric software written to support non-SpeX data.

The 27 observed sources and some of their known properties are listed in Table \ref{targets}.  Columns (1) and (2) list the quasars' J2000.0 coordinates from the SDSS-DR1 catalog.  Column (3) lists the redshift from the SDSS quasar catalog, followed by the SDSS $u$,$g$,$r$,$i$, and $z$ magnitudes in columns (4) - (8).  These are on the $AB$ magnitude system defined by $AB \mathrm{mag} = -2.5 \log_{10} (f_\nu / 3631 \mathrm{Jy})$.  The 2MASS $J$, $H$, and $K$ magnitudes are listed in columns (9), (10), and (11), respectively; these are Vega-based, not $AB$, magnitudes.   About one-third of the objects we observed have detections in the FIRST survey \citep*{Becker95} implying $S_\mathrm{20cm} \geq 1.0$ mJy; columns (12)  and (13) list the 20 cm peak flux density from the FIRST survey (we report the NVSS value \citep{Condon98} for one object which falls outside the FIRST survey area) and the radio-loudness parameter (the ratio of radio to optical flux density $R_r = f_\mathrm{radio}/f_{optical}$) with respect to the $r$ band.  Column (14) reports the absolute $i$-band magnitude, $M_i$.

\section{Constructing the Composite}

When the extraction procedure is applied to the different orders of the cross-dispersed IRTF spectra, it yields spectra with non-linear dispersions.  In the $I$-band order spanning the wavelengths 0.83 - 0.93\micron, the dispersion is about 2.3\AA/pixel and is roughly constant until the $z$-band (0.93 - 1.06\micron) where the dispersion is about 2.7\AA/pixel.  The $J$, $H$, and $K$ orders of the SXD mode have wavelength increments of about 3.2, 4.0, and 5.1\AA/pixel, respectively.  The $K$-band orders (1.9 - 2.55\micron\ and 2.55 - 3.14\micron) in the LXD mode have larger dispersions of about 5.4 and 5.8\AA\ while the two $L$ band orders (2.98 - 3.52\micron\ and 3.52 - 4.2\micron) have wavelength increments of about 6.8 and 8.1\AA/pixel.  These values are relatively constant within the spectral orders, varying by $\pm 0.2$\AA/pixel per order.

To create a mean spectrum it is necessary to shift the objects to their rest frame using the redshifts from the SDSS quasar catalog and to rebin everything to a common dispersion. This causes the rest frame orders to overlap, presenting a challenge for choosing a common dispersion that preserves the most information without oversampling in regions of lower dispersion.  Although the dispersion changes by more than a factor of four over the wavelength range of our spectra, the resolving power, $\mathcal{R} = \lambda/(\Delta \lambda)$, only changes from 3700 to 5000.  Therefore, we chose to rebin our spectra by picking a dispersion that is proportional to the wavelength so that $\mathcal{R}$ is a constant. With this choice, shifting spectra at different redshifts will not change $\mathcal{R}$.  We chose to fix $\mathcal{R}$ to its median value in the observed $K$ band since all our spectra cover this region.  We measure $(\lambda = 21930\mathrm{\AA})/(\Delta \lambda = 5.1\mathrm{\AA}) = 4300$ and build a wavelength grid using this dispersion in $\log \lambda$ increments starting at the shortest rest-frame wavelength in our data, $\lambda_0=5710$\AA.  The recursion relation for the wavelength is

\begin{equation}
\lambda_n = \frac{\lambda_0}{(1-1/4300)^n}.
\end{equation}

The longest rest-frame wavelength in our spectra is 37000\AA, which requires roughly $n=8030$ elements in our wavelength grid.  In this wavelength grid, the dispersion ranges smoothly from about 2\AA\ to 10\AA.  We linearly interpolate the flux and error arrays onto this wavelength grid.

\begin{figure}
\epsscale{1}
\plotone{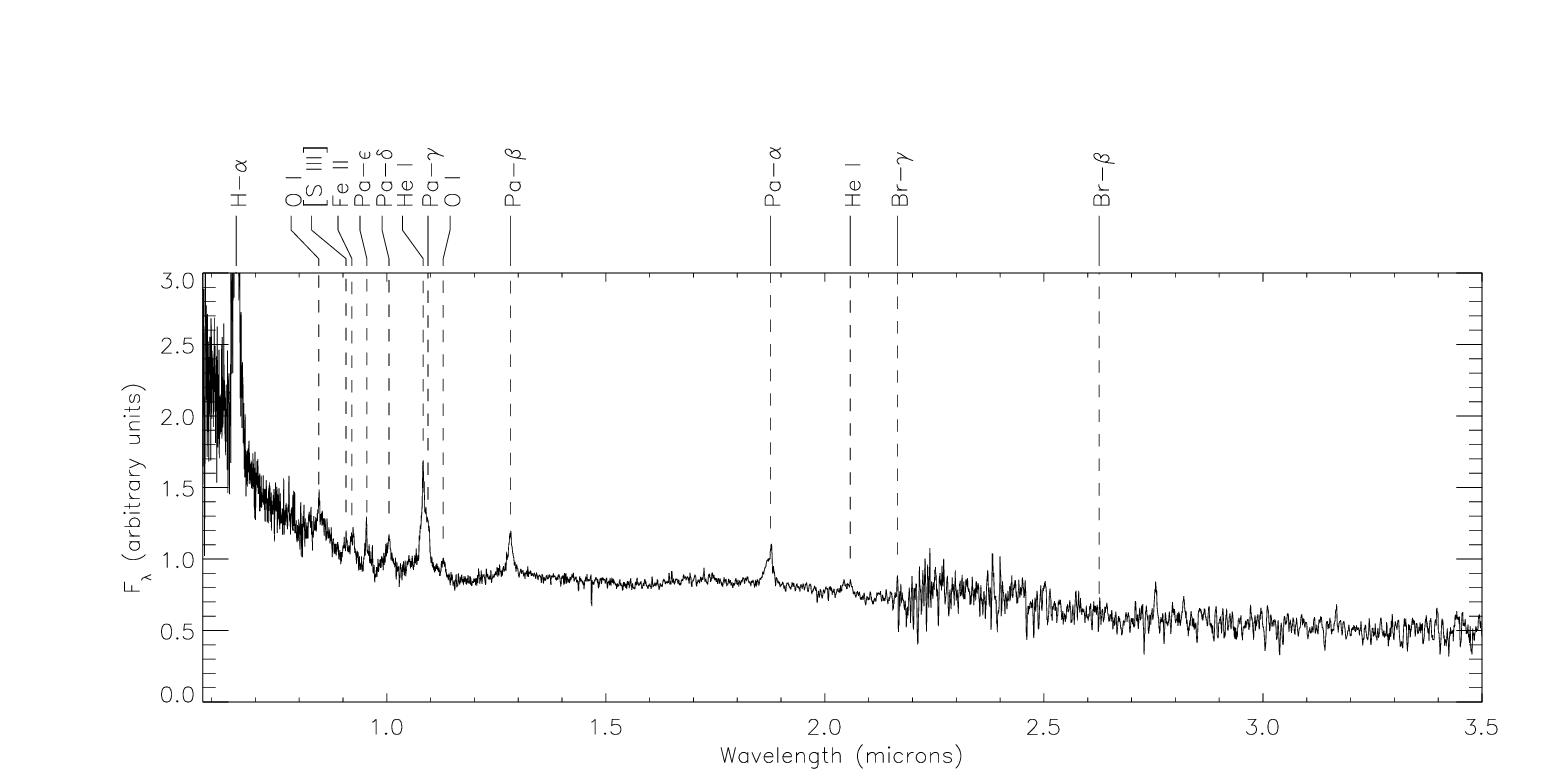}
%\plotone{/nfs/disk/cyrano2/eilatg/composite/spectra/pro/ir_am_comp_wt.ps}
\caption{Near-infrared quasar composite spectrum.  The dotted lines show the positions of expected prominent emission lines: H$\alpha$~6563, O~I~8446, [S~III]~9069, \ion{Fe}{2}~9202, Pa$\epsilon$~9545, Pa$\delta$~10049, He~I~10830, Pa$\gamma$~10941, O~I~11287, Pa$\beta$~12822, Pa$\alpha$~18756, He~I~20520, Br$\gamma$~21654, Br$\beta$~26260 \AA. \label{composite}}
\end{figure}

We built our composite spectrum following the method outlined in \citet{Francis91}.  We ordered the quasars by increasing redshift and scaled the lowest redshift spectrum by its median value, excluding regions of atmospheric absorption.   After removing atmospheric absorption regions and shifting to the rest frame, we then scaled the next spectrum to the first object using the median flux in their overlapping wavelength regions. By using the median flux for scaling each subsequent spectrum, we avoid the influence of strong emission lines.  We weight each spectrum by the inverse square of its error spectrum and sum these products.  We scale subsequent spectra to the median flux in the region overlapping the partial composite from the previous cycle.  The final weighted sum is divided by the total of the weights to yield a weighted average composite spectrum:

\begin{equation}
<f_\lambda> = \frac{\sum_i^n f_{\lambda i}*(1/\sigma_i^2)}{\sum_i^n (1/\sigma_i^2)}.
\end{equation}
Finally, we smooth the resulting spectrum with a boxcar, averaging over 5 pixels.  Figure \ref{composite} displays the final composite with the expected prominent emission lines highlighted by dotted lines.   
 
\begin{figure}
\epsscale{1}
\plotone{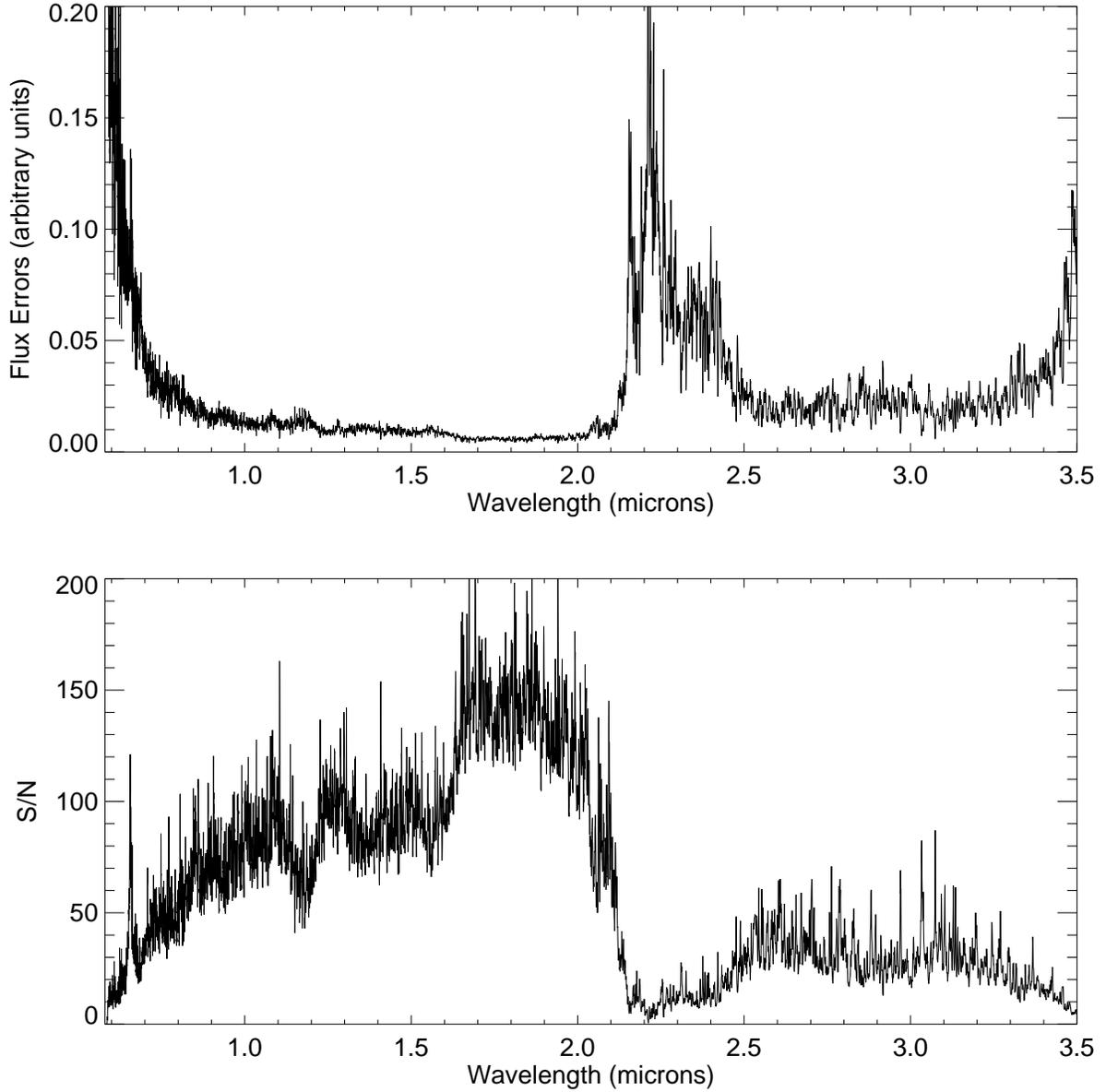}
\caption{Top -- Combined error spectrum. Bottom --Signal-to-noise per pixel of our composite quasar spectrum.\label{error_snr}}
\end{figure}

The upper panel of Figure \ref{error_snr} displays the corresponding combined error spectrum (on an expanded y-axis) obtained by propagating the individual error spectra according to the equation, 

\begin{equation}
\sigma_{tot} = \sqrt{\frac{1}{\sum_i^n (1/\sigma_i^2)}},
\end{equation}
and smoothing in the same manner as for the composite spectrum.  We compute the signal-to-noise ratio per pixel by dividing the composite by the combined error spectrum.  The lower panel of Figure \ref{error_snr} shows the signal-to-noise as a function of wavelength.  We achieve a signal-to-noise ratio of $\sim 150$ in the 1.6-1.9 \micron\ wavelength region, which, in most of our objects, corresponds to the observed $K$-band.  This is where the most photons were collected as a consequence of the overlap of the two SpeX modes in this region.  Spectra covering the observed 0.8 - 4.1\micron\ range have had the overlapping $K$-band region observed twice; the duplicate observations were combined to create the final object spectrum.

\begin{figure}
\epsscale{1}
\plotone{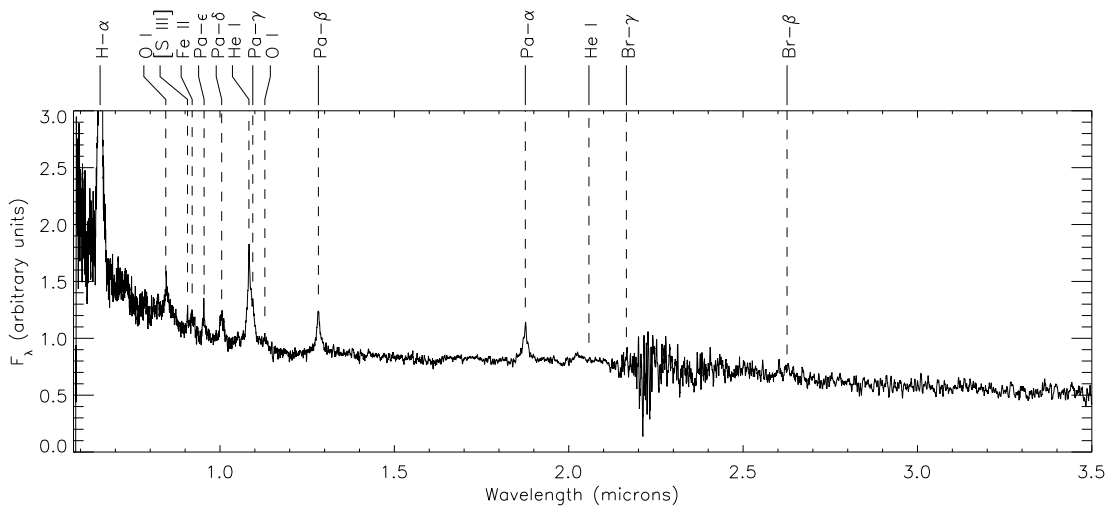}
%\plotone{/nfs/disk/cyrano2/eilatg/composite/spectra/pro/ir_gm_comp_wt.ps}
\caption{Near-infrared quasar composite spectrum, generated from the geometric mean of the individual spectra.  The dotted lines show the positions of expected prominent emission lines: H$\alpha$~6563, O~I~8446, [S~III]~9069, \ion{Fe}{2}~9202, Pa$\epsilon$~9545, Pa$\delta$~10049, He~I~10830, Pa$\gamma$~10941, O~I~11287, Pa$\beta$~12822,Pa$\alpha$~18756, He~I~20520, Br$\gamma$~21654, Br$\beta$~26260 \AA. \label{gmcomposite}}
\end{figure}

\begin{figure}
\epsscale{1}
\plotone{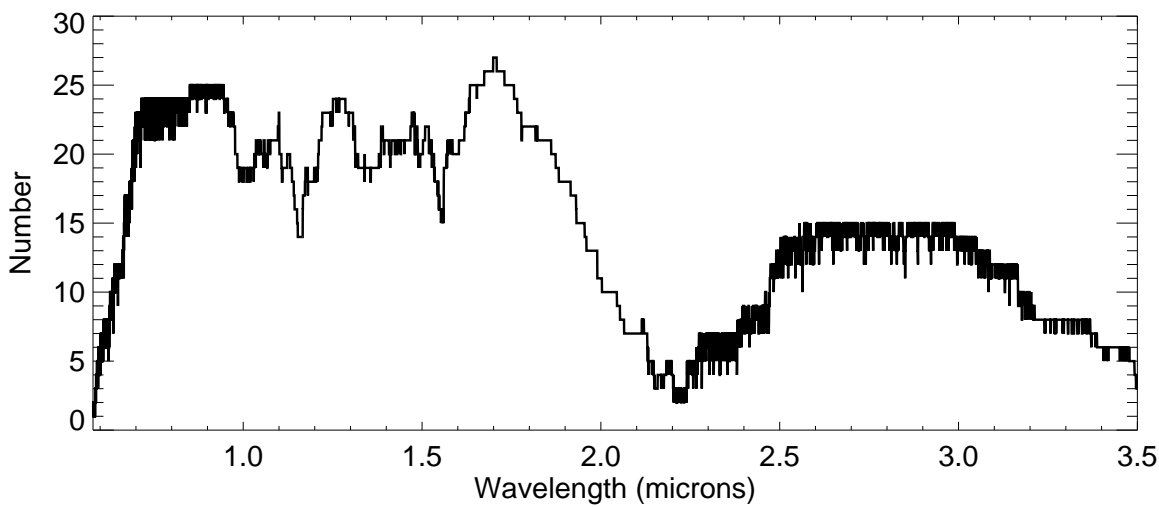}
%\plotone{/nfs/disk/cyrano2/eilatg/composite/spectra/pro/n_contributed.ps}
\caption{Number of spectra contributing to the final composite as a function of wavelength.\label{nspec}}
\end{figure}

Assuming the quasar continuum is a power-law, $f_\lambda \propto \lambda^{-(\alpha_\nu +2)}$, a geometric mean spectrum enables a power-law fit to the continuum that is representative of the arithmetic mean of the indices of the individual spectra, $<f_\lambda>_{\mathrm{gm}} \propto \lambda^{-(<\alpha_\nu> +2)}$.  We have also generated a geometric mean spectrum, $<f_\lambda>_{\mathrm{gm}}$, which does not apply a weighting from the error spectra, according to the following formalism:
\begin{equation}
<f_\lambda>_{\mathrm{gm}} = (\prod_i^n f_{\lambda i})^{1/n},
\end{equation}
where $n$ is the number of spectra contributing to the $i$th wavelength unit.  We scale the spectra using the same scaling as we did for the weighted arithmetic mean, excluding regions of atmospheric absorption in each spectrum.  Figure \ref{gmcomposite} displays the geometric mean spectrum marked with the same emission lines as in Figure \ref{composite}.  Figure \ref{nspec} plots the number of spectra contributing at each wavelength to the geometric mean spectrum.  Table \ref{spectbl} lists the arithmetic mean spectrum, its error spectrum, and the geometric mean spectrum as a function of wavelength.

The order in which spectra are added affects the final composite spectrum.  This was noted by \citet{Brotherton01} who found a change in the spectral index of their composite from $\alpha = -0.46$ ($F_\nu \propto \nu^\alpha$) when combining the spectra starting from the lowest redshift to $\alpha = -0.49$ when starting with the highest redshift quasars.  The first few spectra in the composite-building algorithm can strongly influence the shape of the final composite if they contain wavelength regions that exist in only a few spectra.  \citet{Telfer02} order their spectra beginning with objects that contribute to the best-sampled part of their wavelength coverage.  We constructed two additional composite quasar spectra to study the dependence of the spectral shape of our composite on the order in which we combine the spectra.  Following \citet{Brotherton01}, we made one spectrum starting with the highest redshift quasar and adding sequently lower redshift objects to the composite. To make a composite using the method of \citet{Telfer02}, we detemined (as shown in Figure \ref{error_snr}) that quasars within the redshift range $0.188 < z < 0.316$ contribute to the $1.6-1.9$ \micron\ wavelength region, which has the highest signal-to-noise in the composite.  We started with SDSSJ015910.0+010514.5, a $z=0.2174$ quasar that has both SXD and LXD coverage and added to it spectra that have SXD and LXD coverage in same wavelength range.  We then added spectra in that wavelength range having only an SXD or LXD spectrum.  Finally, we added sources with redshifts outside this high-SNR range, alternating between lower and higher redshift objects.  We discuss the sensitivity of the resultant spectra to these different ordering methods in \S 4.  The analysis of our near-infrared quasar composite described below pertains to the composite made from quasars added in increasing redshift order.

We also constructed a uniformly weighted composite spectrum in order to verify that our weighting scheme does not skew our composite toward the highest signal-to-noise spectra.  In this case, instead of using equation (2) in our composite-building algorithm, we use $<f_\lambda> = \sum_i^n f_{\lambda i}/n$, excluding the same regions of atmospheric absorption as in the error-weighted composite.  The uniformly weighted spectrum has a similar continuum shape and line profiles as the error-weighted composite.  We verified this by examining the ratio of, and difference between, the two spectra.  No obvious feature emerged and there were no residuals at the location of emission lines.  

The signal-to-noise ratios in the observed $K$-band for the individual spectra that make up our composite range from $\sim 1.3 - 60$; the median signal-to-noise ratio is 12.4, the mean is 16, and fourteen of the 27 quasars in our sample have signal-to-noise ratios between 10 and 30.  Comparison of our uniformly weighted composite and the highest signal-to-noise spectrum shows strong residuals both in the continuum and at the position of emission lines.  We conclude that our error-weighted composite is not dominated by the highest signal-to-noise spectra in our sample.

\section{Analysis of the Composite Spectrum}

\subsection{Is Our Composite Biased?}

\begin{figure}
\epsscale{1}
\plotone{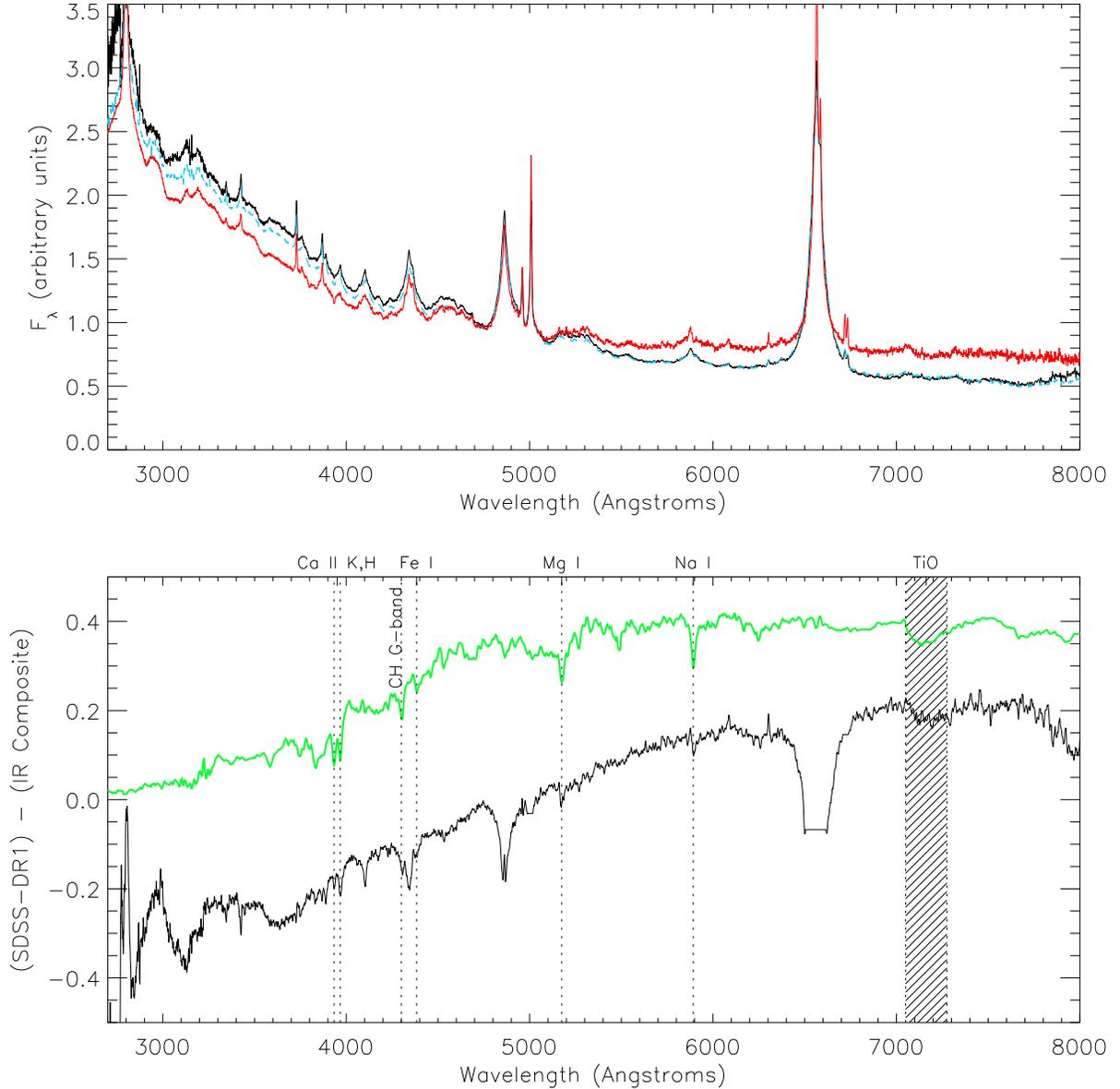}
%\plotone{/nfs/disk/cyrano2/eilatg/composite/spectra/pro/sdss_opt_comp_compare.ps}
\caption{Upper -- Optical composite spectrum generated from the 27 objects in the near-infrared sample ({\it black}).  Overplotted are composites which we generated from the 87 SDSS-DR1 quasars that obeyed our selection criteria ({\it blue}), and the SDSS quasar composite spectrum from Vanden Berk et al. (2001) ({\it red}).  Lower -- Difference between the optical composite from this work (black line above) and the SDSS-DR1 composite spectrum (red line above) ({\it black}). The template spectrum of a giant elliptical galaxy from \citet{Mannucci01} is overplotted (with an arbitrary scaling, {\it green}) for comparison.  The residual spectrum reveals both continuum and absorption features obvious in the galaxy spectrum.  The absorption lines marked with a {\it dotted line} are \ion{Ca}{2} K 3933, \ion{Ca}{2} H 3968, CH G-band 4300, \ion{Fe}{1} 4384, \ion{Mg}{1} 5175, \ion{Na}{1} 5892 \AA.  The prominent TiO absorption between 7050-7275\AA\ is also marked by the shaded band. \label{sdss_compare}}
\end{figure}

To evaluate any bias our selection criteria may introduce with respect to the parent quasar sample, we constructed an optical composite spectrum using the technique described above from the 27 SDSS optical spectra available for our sample.  We rebinned the spectra using the same resolving power, $\mathcal{R}=4300$, and applied the same scaling technique after sorting the spectra by increasing redshift.  We also produced optical composites using the two additional sorting methods: we sorted the quasars by decreasing redshift and sorted according to the objects whose redshifts contribute to the highest signal-to-noise part of the final composite in the same manner described in \S3.  The SDSS database provides an error spectrum for each object which we use to weight the data in the same manner as for the near-infrared composite.  The top panel of Figure \ref{sdss_compare} plots the resultant optical composite ({\it black}) for our 27 objects along with a composite that we generated from the complete sample of 87 SDSS-DR1 quasars that obeyed our selection criteria ({\it blue}). We compare these with the SDSS quasar composite spectrum created from the parent DR1 sample by \citet{VandenBerk01} ({\it red}).  Our selection criteria do introduce a small bias which leads to a quasar with a {\it bluer} continuum than the total SDSS sample composite. 

\citet{VandenBerk01} fit two power laws to the quasar continuum, fitting one spectral index for wavelengths shorter than 4000 \AA\ and one for wavelengths longer than 4000 \AA.  To compute the spectral index for wavelengths longer than 4000 \AA, \citet{VandenBerk01} fit a power law to the line-free wavelength ranges 6006-6035 \AA\ and 7160-7180 \AA.  They find a spectral index of $\alpha_\nu = -2.45$ using their geometric mean spectrum and $\alpha_\nu = -1.58$ using their median-combined spectrum.  We fit a power law to the same wavelength regions in our optical composite and measure a spectral index of $\alpha_\nu = -1.03\pm0.1$. In Section 4.2 we analyze the continuum properties of our near-infrared composite spectrum which also measures the spectral index around H$\alpha$.  These measurements are shown in Table \ref{spec_index}.  The wavelength regions used by \citet{VandenBerk01} to fit the blue end of their composite (1350-1365 and 4200-4230 \AA) are not available using our composite made of quasars with $z\leq 0.5$.  However it is clear from Figure \ref{sdss_compare} that the SDSS composite has a softer UV spectrum than that produced from our subsample.  

This discrepancy is likely a consequence of the varying properties of quasars that contribute to the blue and red ends of the SDSS composite spectrum\footnote{We find no distinction in the red to near-infrared ($i-K_s$) colors between our sample and the SDSS-DR1 objects with the same redshift limit as our sample ($z<0.5$). The median $i-K_s$ colors for our sample and the SDSS-DR1 quasars with $z<0.5$ are 3.1 and 3.2, respectively.}.  As the composite moves to longer rest-wavelengths, more lower-$z$ objects with lower average luminosities contribute.  The median absolute $i$ magnitude, $M_i$, for our sample is $-23.7$ and our maximum redshift is $z=0.418$. In contrast, the objects contributing to the blue end of the SDSS composite spectrum can have redshifts as high as $z\sim 1.7$, and the median $M_i$ for objects with $z\lesssim 1.7$ is $-25.0$.  At the red end of the SDSS spectrum, however, only objects with $z\lesssim 0.4$ contribute; their median $M_i$ is $-22.5$, fainter than our requirement of $M_i < -23.0$.  

The difference spectrum is illustrated in the lower panel of Figure \ref{sdss_compare}.  For comparison, we overplot the elliptical galaxy template spectrum from \citet{Mannucci01} ({\it green}) to determine if any galaxy features are present in the residuals.  The result strongly suggests that most if not all of the excess red emission in the SDSS quasar composite spectrum is due to host galaxy light.  We identify with {\it dotted lines} obvious continuum and absorption line features which appear in our residual spectrum as well as the galaxy template.  We shade the prominent TiO absorption band which appears in both spectra between 7050\AA\ and 7275\AA.  We conclude that the inclusion of low-redshift, low-luminosity AGN in the SDSS composite on the red end of the spectrum adds host galaxy light which enhances the red continuum.  The sources in our sample were selected to be luminous enough that host galaxy light would not be a significant factor in the continuum.  In addition, our slit width (0\farcs5 in the SXD mode) takes in only half as much galaxy light as the SDSS spectroscopic fibers (3\arcsec) assuming an elliptical host with a de Vaucouleurs ($r^{1/4}$-law) surface-density profile and a half-light radius of 11.5 kpc.

Furthermore, in a comparison between SDSS and Palomar-Green quasars from the Bright Quasar Survey \citep[BQS][]{Schmidt83} \citet{Jester05} found that quasar samples with brighter apparent magnitudes (in $B$, $g$ and $i$ bands) are biased to the blue compared with a fainter sample (see their \S4.1.3, and Figure 10).  Our sample also selects bright objects because of our $K_s < 14.5$ requirement.  In the SDSS-DR1 catalog, the median $g$ and $i$ magnitudes for quasars with $z<0.5$ and $M_i < -23.0$ but no magnitude restriction are 18.33 and 18.98, respectively.  By imposing the $K_s < 14.5$ magnitude limit, the median $g$ and $i$ magnitude in our sample are 17.27 and 16.96, respectively.  Therefore the bluer continuum of our composite compared with the SDSS composite is also due to the color bias introduced by our bright magnitude criterion.

\subsection{Continuum}

\begin{figure}
\epsscale{1}
\plotone{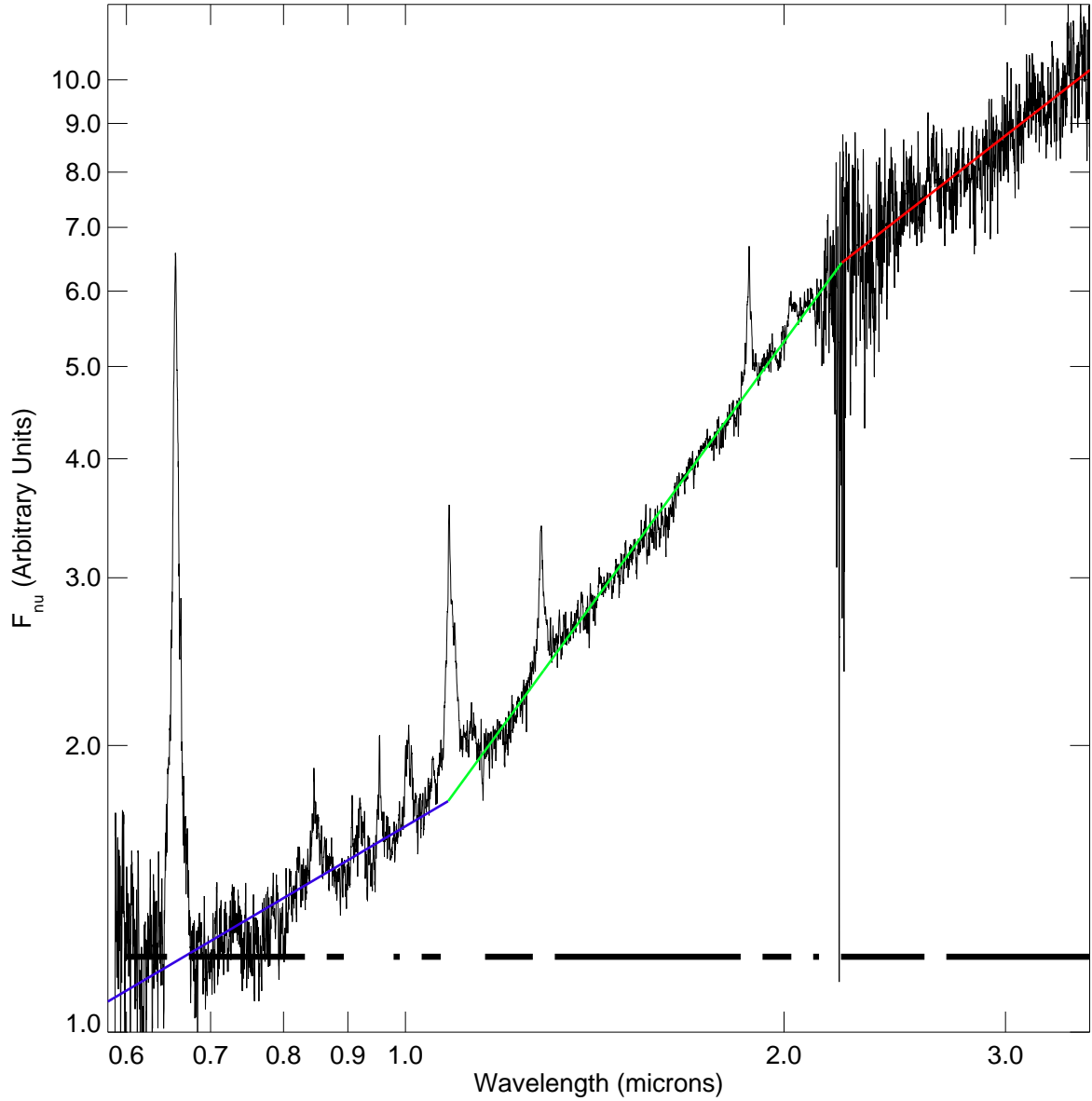}
%\plotone{/nfs/disk/cyrano2/eilatg/composite/spectra/pro/sp_index_fit.ps}
\caption{Composite quasar spectrum plotted in $F_\nu-\lambda$ on logarithmic axes with power-law fits to three spectral regions overplotted in {\em blue} ($0.58-0.97$ \micron), {\em green} ($0.97-2.37$ \micron) and {\em red} ($2.37-3.50$ \micron). The thick black lines at the bottom mark the continuum windows used in the power-law fit. \label{sp_index}}
\end{figure}

A single power law does not fit the continuum from 0.58-3.5\micron\ well.  This is evident in Figure \ref{sp_index} where we display the composite in $\lambda~vs.~F_\nu$ space on logarithmic axes.  We fit a broken power-law (with two breaks) to the spectrum using the IDL fitting routine AMOEBA, a robust fitting algorithm that performs a $\chi^2$ minimization for a given model and data while ignoring outliers;  the algorithm does not return errors on the fitted parameters.  However, AMOEBA does require a fitting tolerance parameter, which we specify to be $1.0\times10^{-3}$.  The broken power law model has six parameters: the two break points in the power law, the amplitude of the function at the first break and the three power law indices, one for each segment.  We exclude Balmer, Paschen and Brackett lines as well as major ionic lines (see \S 4.3) from the fit.  The wavelength windows used in our fit are marked with a thick line at the bottom of Figure \ref{sp_index}.

Table \ref{spec_index} compares the spectral indices derived for different regions of our spectra.  For the composite generated by sorting the spectra from lowest to highest redshifts (row 6), we find that the optical part of the spectrum containing H$\alpha$ is best fit by a power law starting at 5700 \AA\ and breaking at 10850 \AA\ with a spectral index of $\alpha=-0.78$ (where F$_\nu \propto \nu^\alpha$).  This spectral index is considerably harder than that found by \citep{VandenBerk01} in the $6005-7180$ \AA\ region ($\alpha_{\nu\mathrm{SDSS}} = -2.45$), although it is consistent with the $6005-7180$ \AA\ slope of $-1.03$ derived for our optical composite (row 3).  The $1.085-2.230$ \micron\ region is best fit with an index of $\alpha=-1.81$, while the 2.23-3.5 \micron\ region is fit by a harder spectral index of $\alpha=-1.03$.

It is clear from Table \ref{spec_index} that spectral indices measured in the UV-optical part of composite quasar spectra agree \citep[see also Table 5 in][]{VandenBerk01}.  The order in which spectra are combined does not have a strong affect on the spectral slope in the 1-2 \micron\ and 2-3.5 \micron\ regions.  The largest discrepancies occur in the red-optical region around H$\alpha$.  Here, the order of combining the spectra introduces a variation of up to $\Delta\alpha_\nu \sim 0.5$. 

We also note that the second break in the spectrum occurs in the noisiest part of the composite where the LXD order begins to contribute.  This parameter is also sensitive to the order in which the spectra are combined, shifting the break by up to $\Delta\lambda \sim 0.3$ \micron.  The two slopes are very different and a break must occur between the two regions; however, the location of the break may shift when more spectra can be added to the composite to improve its signal-to-noise ratio.  

\begin{figure}
\epsscale{1}
\plotone{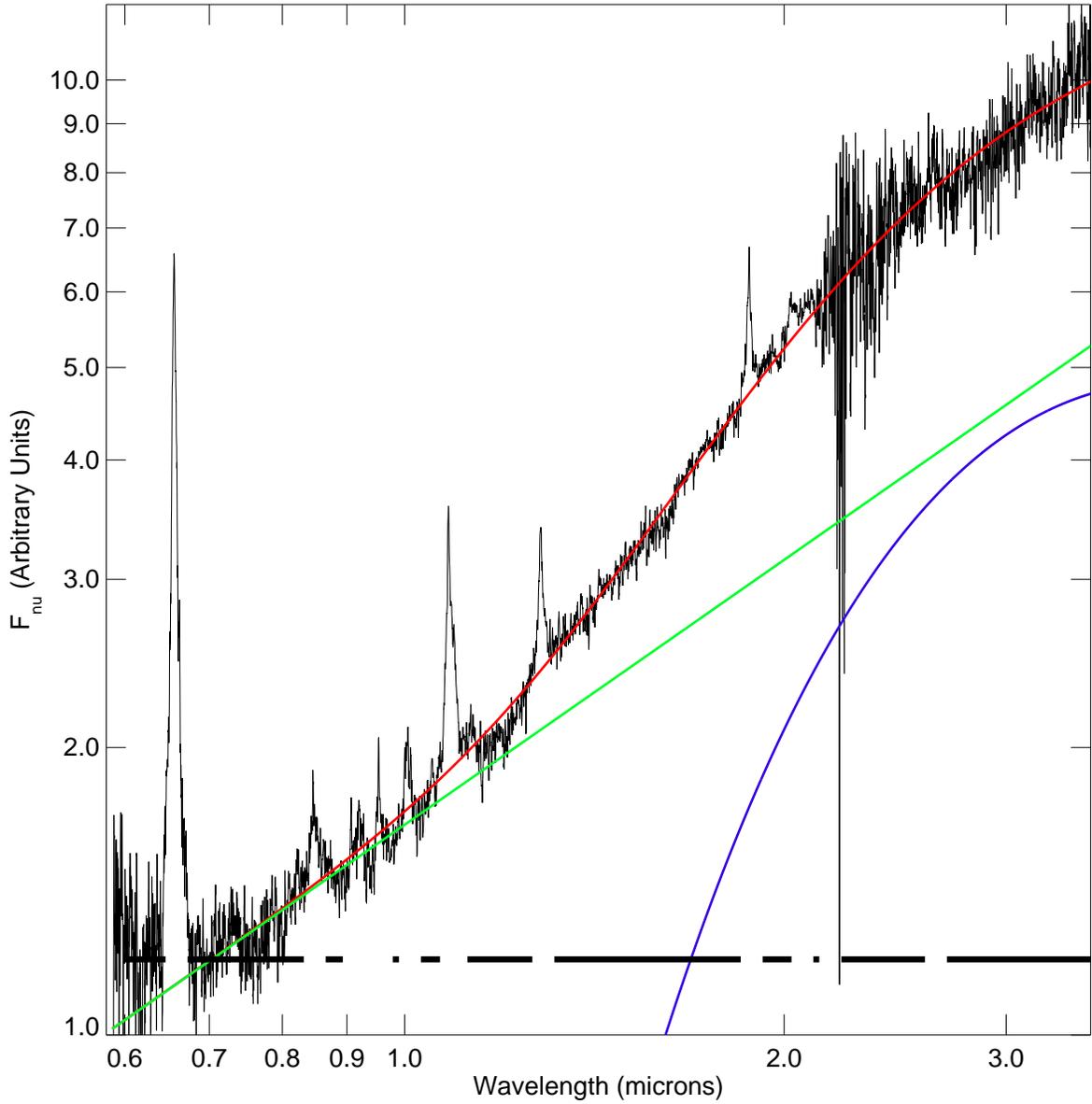}
%\plotone{/nfs/disk/cyrano2/eilatg/composite/spectra/pro/hotdust_fit.ps}
\caption{Power-law plus blackbody fit to our composite quasar spectrum plotted in $F_\nu-\lambda$ on logarithmic axes.  The combined fit is shown in {\em red}, the power law component with a best-fit spectral index of $\alpha = -0.92$ is shown in {\em green} and the blackbody with the best-fit dust temperature of $T_d=1260$ K is shown in {\em blue}. \label{hotdust}}
\end{figure}

Excess emission between $\sim 2$ and 10 \micron\ has been widely noted in previous work and has been attributed to hot dust at $\sim 1500$ K in some studies of AGN spectral energy distributions (SEDs) \citep{Rieke78,Edelson86,Barvainis87,Elvis94}.  We fit a power-law plus a blackbody to our composite spectrum with AMOEBA.  We define the following function,
\begin{equation}
F_\nu = C_1 \nu^\alpha + C_2 B_\nu(T_{dust}),
\end{equation}
where $B_\nu(T_d)$ is the Planck function, allowing AMOEBA to fit $C_1$, $\alpha$, $C_2$, and $T_{dust}$.  We initialize the scaling factors to unity and the spectral slope, $\alpha$, to $-0.78$, the value for the optical portion of the spectrum obtained from our broken power-law fits, and the dust temperature, $T_{dust}$, to 1500 K as found by \citet{Barvainis87}.  The fit converged on a spectral slope of $-0.92$ and a temperature of 1260 K.  We list the result of this fit in Table \ref{bb} along with fits to composites generated by ordering the spectra in the different schemes described in \S 3.  The order in which the spectra are combined in generating the composites does not effect the fitted temperature or slope by more than 10\degr K or $\Delta\alpha_\nu \sim 0.02$. Figure \ref{hotdust} shows the results of our fit to the composite ordered by increasing redshift.  This model reproduces the continuum shape at least as well as the broken power-law fits (fits to both models yield $\chi^2 \sim 3$) and uses two fewer parameters and a stronger physical motivation. This strongly suggests a hot dust component exists in the luminous blue quasars observed in our sample.

The detection of silicate emission at 10 and 18 \micron\ in {\em Spitzer} spectra of PG quasars also suggests dust, though at lower temperatures, ranging from 140 K to 220 K \citep{Hao05}. This is not inconsistent with our result, since different dust temperatures are expected at different radial distances from the AGN. \citet{Barvainis87} fit a more complex model including a hot dust component over a range of radii to the SEDs of 3C273 and IRAS 13349+2438 and found temperatures ranging from $\sim 100$ K to 1500 K.  The wavelength coverage of our composite spectrum allows us to probe the hottest grains in this range.  

\subsection{Emission lines}

We manually identified emission lines in the composite spectrum and measured their intensities by fitting a Voigt profile to each isolated line.  The Voigt fitting algorithm computes a linear continuum (2 parameters) as well as a total intensity, central peak position, doppler width and damping width (4 parameters).  We allowed the peak position in the fit to vary.  We integrated the continuum component of the fit and subtracted it from the integrated intensity over the wavelength region containing at least $5\%$ of the peak fitted flux to compute the intensity of the line above the continuum.  We also computed the equivalent width of each line using the fitted profile.  In the case of the blended \ion{He}{1} + Pa$\gamma$ pair, we fitted a double Voigt profile which would require ten parameters  (2 continuum parameters and two sets of four parameters for the individual Voigt profiles).  Owing to the strong blending of the lines, we fixed their central peak wavelengths to the known laboratory values of 10830\AA\ and 10941\AA, respectively.  We then computed the intensity of each line with a single Voigt profile using the six fitted parameters pertaining to each line, also over the region containing $\geq 5\%$ of the fitted peak flux.  

The low signal-to-noise of some of the weaker lines in our spectrum made the Voigt fitting algorithm a better choice than a multiple-component gaussian fit. Single gaussian profiles do not model well the broad wings of the emsission lines and generate a poor fit.  Additional fitting components account for these wings.  However, in poor signal-to-noise lines (such as [\ion{S}{3}] $\lambda$ 9069 \AA\ and \ion{Fe}{2} $\lambda$ 9202 \AA) the additional gaussian components will fit the noise and the fitted lineshape will appear distorted.  The fixed shape of the Voigt profile, with its four line parameters provide more realistic fits to our lines, given their range in signal-to-noise.

\begin{figure}
\epsscale{1}
\plotone{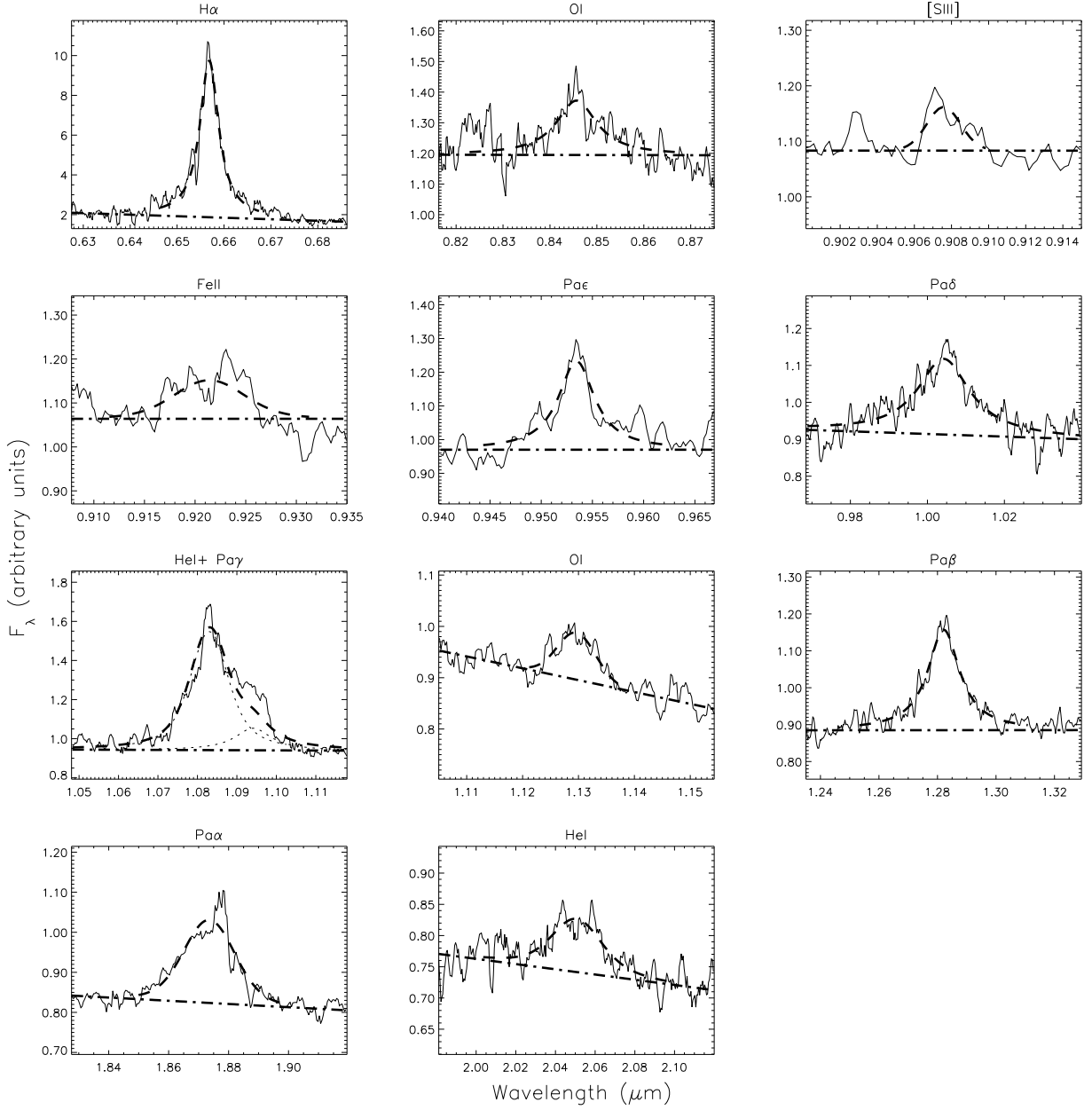}
%\plotone{/nfs/disk/cyrano2/eilatg/composite/spectra/pro/line_fits.ps}
\caption{Fitted emission-lines displayed over their fitted regions.  The fitted profile is overplotted as a {\em dashed line} ({\em red dotted line} in electronic edition) and the continuum is plotted with a {\em dash-dot} line ({\em blue dash-dot} in electronic edition). In the \ion{He}{1}+Pa$\gamma$ panel, we overplot the individual line profiles with a {\em dotted line}. \label{linefig}}
\end{figure}

Figure \ref{linefig} displays the emission lines over their fitted regions.  The line profile is plotted as a {\em dotted line} over the region used to integrate the intensity and the continuum is plotted in a {\em dash-dot line}. Table \ref{linetbl} lists the line parameters extracted from the fits.  Column (1) identifies the feature.  Column (2) lists the laboratory wavelength for the feature followed by the fitted peak position of the line in column (3) (except in the case of the blended \ion{He}{1} + Pa$\gamma$ pair where we fixed these values).  Columns (4) and (5) list the lower and upper wavelength limits for integrating the line intensity.  Column (6) lists the line intensity relative to H-$\alpha$ followed by the equivalent width of the line in column (7).  Column (8) lists the full-width at half-maximum of the line.  

The relative fluxes of Pa$\alpha$ and Pa$\beta$ give a line ratio of Pa$\alpha$/Pa$\beta$ = $0.64\pm0.01$.  This value is less than half that predicted by Case B recombination which gives a value of 1.8 for a typical broad line cloud with $T_e = 10^4$ K and $N_e = 10^9$ cm$^{-3}$. The expected value is constant to within $\pm 0.1$ for temperatures ranging between $5\times 10^3$ K and $3\times 10^4$ K and densities ranging between $10^8$ cm$^{-3}$ and $10^{10}$ cm$^{-3}$ \citep{Hummer87}.  Our measurement is consistent with that of \citet{Soifer04} who obtained a {\em Spitzer} IRS spectrum of the $z=3.91$ quasar, APM08279+5255, and found Pa$\alpha$/Pa$\beta$ = $1.05\pm0.2$ which is also too small for Case B recombination.

\section{Radio-loud versus Radio-quiet subsamples}

Twenty-two of the twenty-seven quasars in our sample are covered by the FIRST radio survey \citep{Becker95}.  Eight are detected, with 20 cm flux densities ranging from 1.3 to 166 mJy; the remaining sources have upper limits $S_{20}<1.0$ mJy.  The five objects falling outside the FIRST survey area are covered by NVSS \citep{Condon98}; one is detected with a flux density of 2.4 mJy (the remainder have $S_{20}$ below this value).  We compute the radio loudness parameter for each source starting with the formalism of \citet{Ivezic02} which defines $R_m = \log(F_\mathrm{radio}/F_\mathrm{optical})=0.4(m-t)$, where $m$ is any of the SDSS magnitudes, and $t$ is the AB radio ``magnitude'' of the FIRST flux density, provided in the DR1 catalog from which we selected our sources.  We use the radio-loudness parameter  without the logarithm (i.e., $10^{R_m}$) and report its value in column (13) of Table \ref{targets}. All have $R\gtrsim 1.0$.  We do not include a $K$-correction when computing $R_r$ due to the unknown radio spectral indices for these quasars.  Instead, we examined $R_m$ for all five SDSS filters and found that in no object did $R_m$ range over more than a factor of $\sim2$.  In our analysis we use $R_r$ from the SDSS $r$ band.

Definitions for what constitutes a ``radio-loud'' quasar range from $R > 10$ to $R > 100$ while radio-quiet quasars have $R\lesssim 1.0$.  Depending on the definition, our sample contains between two and five radio-loud objects.  The rest of the sample has $R<10$ and are either considered radio-intermediate or radio-quiet. Although we have only a small sample of radio-loud objects, we can compare radio-loud and radio-quiet composites to look for any obvious differences.
 
\begin{figure}
\epsscale{1}
\plotone{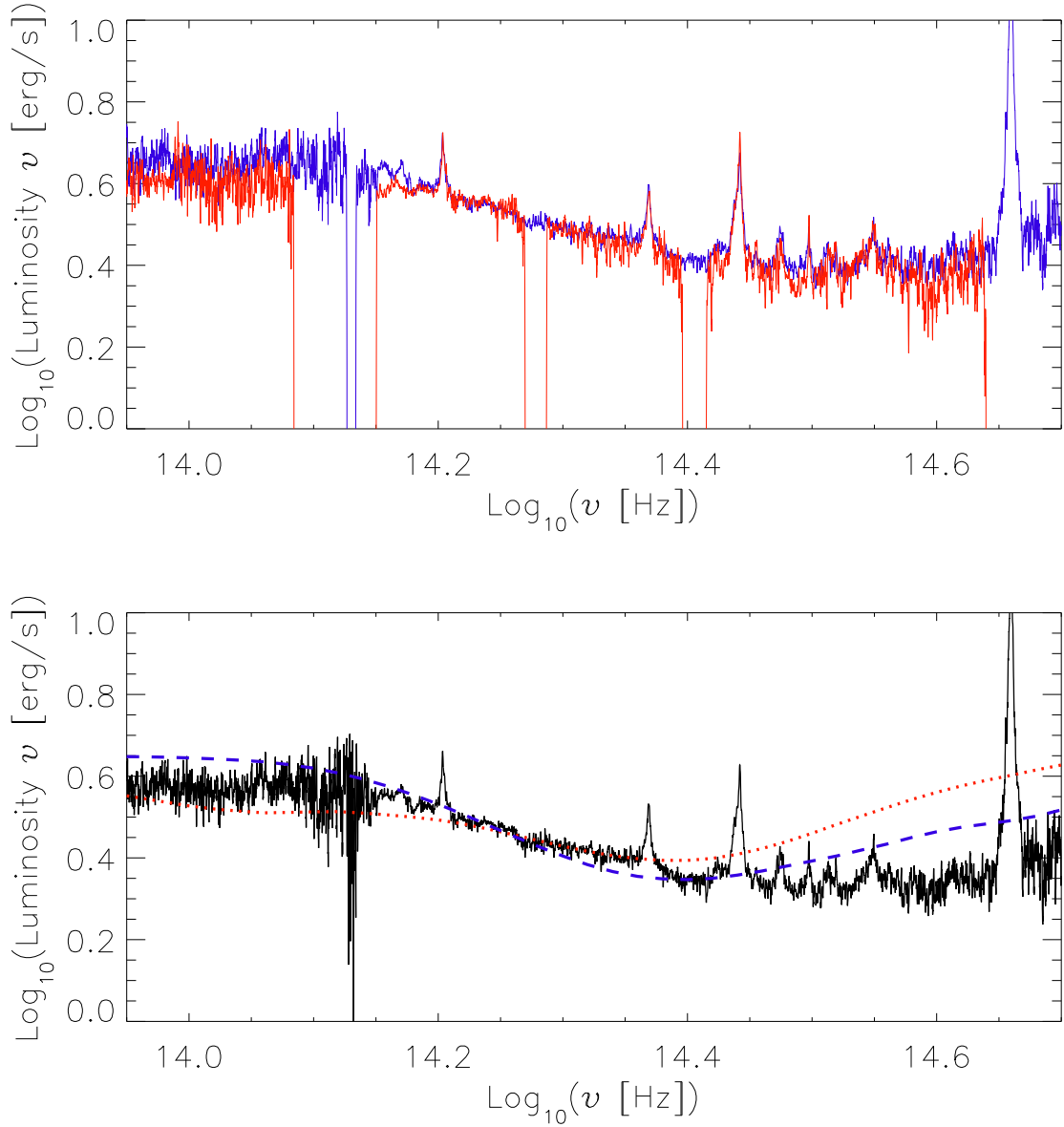}
\caption{Top -- Comparison of composites made from radio-quiet ({\em blue}) and radio-loud ({\em red}) subsamples plotted in $\log(\nu)~vs.~\log(\nu \mathrm{L_\nu})$. Bottom -- The geometric mean composite spectrum from the entire sample overplotted with the mean spectral energy distributions from Elvis et al. (1994) for radio-loud ({\em red dotted line}) and radio-quiet ({\em blue dashed line}) quasars.\label{rl_rq}}
\end{figure}

We have created two composites from the radio-loud (including all five objects with $R>10.0$) and radio-quiet (including all other spectra) subsamples.  The top panel of Figure \ref{rl_rq} plots the radio-quiet composite in {\em blue} with the radio-loud composite overplotted in {\em red}.  We plot these in $\log(\mathrm{Luminosity}\times \nu) \ vs. \ \log(\nu)$ space which is standard for plotting spectral energy distributions.  The radio-loud spectrum is scaled to the radio-quiet spectrum at $\nu=1.8\times 10^{14}$ Hz, or 1.67\micron, roughly where the signal-to-noise of our composite is highest. 

%We fit broken power-laws to the radio-loud and radio-quiet spectra using the method described in \S 4.2 to obtain spectral indices and the location of the breaks.  The resultant parameters are reported in rows 9 and 10 of Table \ref{spec_index}.  The radio loud-spectrum has a smaller wavelength span, starting at $\sim 7000$ \AA,  and has several gaps in its spectrum because of the small number of quasars used in its construction.  There is strong discrepancy between the two spectra at the location of the short-wavelength break and the spectral indices blueward of this break. This is likely caused by the lack of data below $\sim 7000$ \AA\ in the radio-loud composite.  The power-law indices in the middle of the spectrum are also quite different ($-1.65$ for the radio-quiet spectrum and $-1.96$ for the radio-loud spectrum), although the long wavelength break points are consistent ($\sim 1.8$ \micron), and the spectral shapes appear very similar.  We caution that differences between the radio-loud and radio-quiet spectra may result from the small sample size.  

Superficially the spectra appear to show only small differences. This is likely a result of the small number of radio-loud quasars and our liberal definition of ``radio-loud'' ($R_r \geq 10.0$); a larger sample of radio-loud quasars, with $R_r \geq 100$ may reveal significant differences between the spectra.   

We compare our quasar composite spectrum made from the entire sample with the SED for quasars derived by \citet{Elvis94} in the lower panel of Figure \ref{rl_rq}. For consistency the SEDs are scaled to our quasar composite at the same wavelength used in the top panel.  The radio-loud SED is plotted with a {\em red dotted line} and the radio-quiet SED is displayed with a {\em blue dashed line}.  Some of the differences observed between the radio-quiet and radio-loud subsamples in the top panel are present in the SEDs of the bottom panel, particularly below $\sim 1.6\times10^{14}$ Hz ($\log(\nu) < 14.2$).  The radio-loud quasars have a harder spectrum in the $\sim 1.2-2$ \micron\ band.  Spectra obtained at longer wavelengths will determine whether this trend continues into the mid-infrared.  In the SEDs of \citet{Elvis94}, when normalized in the same way as in Figure \ref{rl_rq}, the trend reverses around $10^{13}$ Hz, below which the radio-loud continuum gains strength out into the radio regime, as expected.  Above the second break in the spectra, around $2.5\times 10^{14}$ Hz ($\log(\nu) > 14.4$), the radio-quiet and radio-loud SEDs from \citet{Elvis94} have harder continua than our composite continuum.  The difference may be a consequence of the higher median luminosities of the quasars chosen for the \citet{Elvis94} SED.  Their radio-loud SED has 18 radio-loud quasars with a median $M_V = -25.3$ and 29 radio-quiet quasars with a median $M_V = -23.8$.  The median luminosity of our sources, in the equivalent SDSS $g$-band is $M_g = -23.4$. 

We also fit the power-law plus blackbody model to our radio-loud and radio-quiet composites.  We report these results in Table \ref{bb}.  The fit to the radio-quiet spectrum yields a softer spectral-index, $\alpha=-1.34$, but a dust temperature consistent with that found for the total composite, $T_{dust} = 1260$ K.  Since the radio-loud composite covers a smaller wavelength span and lacks coverage in the optical portion of the spectrum where the power-law dominates, the fit to equation (5) does not yield physical results ($\alpha= -1.42$, and $T_{dust} = 5$ K).  Instead, we fit both composites to equation (5) fixing either the spectral-index or the dust temperature to the value derived for the total composite, $\alpha = -0.92$ or $T_{dust} = 1260$ K.  This approach allowed for fewer free parameters and generated resonable results, consistent with the total composite.  When one parameter was held fixed, the radio-loud spectrum was fit by either a $T_{dust}$ hotter by 100 K or a slightly softer spectral index compared with the radio-quiet spectrum.  

\section{An Optical-to-Near-Infrared Composite} 

We used our composite-spectrum-building algorithm described in \S3 with the optical spectra provided by SDSS and our IRTF spectra to build a composite quasar spectrum spanning the wavelength range 2700-37000\AA.  We applied the same wavelength binning as in Equation (1) with $\lambda_0=2700$ \AA\ and $n=11250$.  The optical spectra were incorporated into the composite individually as separate spectra.  They were first ordered by increasing redshift as for the infrared counterpart.  Then, since each optical spectrum overlaps its infrared counterpart between roughly 8000\AA\ and 9500\AA, we used this region to scale and incorporate the first optical spectrum into the composite.  After the full spectral range is filled, each additional spectrum (infrared or optical) is scaled to the previous partial composite using the median values of their mutually overlapping regions.  

\begin{figure}
\epsscale{1}
\plotone{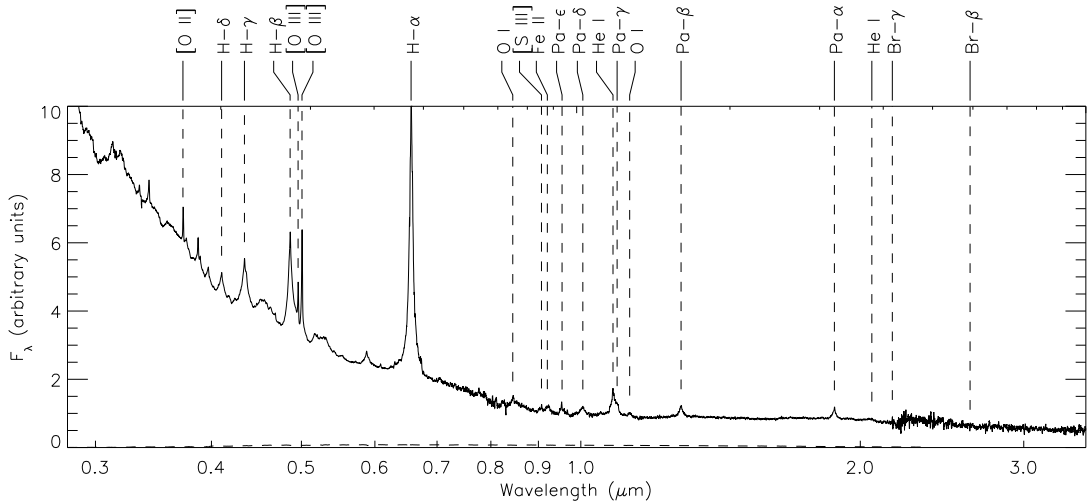}
%\plotone{/nfs/disk/cyrano2/eilatg/composite/spectra/pro/opt_ir_am_comp_wt.ps}
\caption{Optical-to-near-infrared quasar composite spectrum. Expected prominent emission lines include ({\em vertical lines}):[O~II]~3727, H$\delta$~4102, H$\gamma$~4341, H$\beta$~4862, [O~III]~4959, [O~III]~5007, H$\alpha$~6563, O~I~8446, [S~III]~9069, Fe II~9202, Pa$\epsilon$~9545, Pa$\delta$~10049, He~I~10830, Pa$\gamma$~10941, Pa$\beta$~12822, Pa$\alpha$~18756,  Br$\gamma$~21654, Br$\beta$~26260 \AA.  The {\em dashed line} shows the possible contribution from a $2\times L^\star$ elliptical host galalxy. \label{optircomposite}}
\end{figure}

Figure \ref{optircomposite} shows the resulting spectrum plotted as log-wavelength versus flux.  We also plot a model host-galaxy composite spectrum (dashed line) to represent an estimated host galaxy contribution to our quasar composite.  We generated the galaxy spectrum according to the following prescription:  We assume that each quasar in our sample is hosted by a massive elliptical with a luminosity of $2 \times L_{K_s}^\star$.  This is supported by observations of nearby quasar hosts \citep[cf.][]{Percival01,Dunlop03,Floyd04}. We use $M^\star_{K_s} = -23.47$ for early type galaxies \citep{Kochanek01} and scale the elliptical galaxy template spectrum from \citet{Mannucci01}, which spans ultraviolet to near-infrared wavelengths, by its $K$-band flux to the flux expected from an $2 \times L^\star$ galaxy at the redshift of each quasar.  We then scale the final template by $20\%$ to account for the fraction of galaxy light entering through our slit.  At its maximum, the putative host galaxy contributes $\sim 6\%$ of the overall continuum around 0.9-1.0 \micron\ (cf. \S2).  Table \ref{optirspectbl} lists the arithmetic and geometric mean spectra, available in their entirety in the electronic edition of the Journal.

\subsection{K-corrections and Infrared Colors}

A particularly useful application for this quasar composite is in calculating $K$-corrections to passbands that lie in our spectral region.  These include the three 2MASS bands, $J$ (1.25 \micron), $H$ (1.65 \micron), and $K_s$ (2.17 \micron), as well as the shortest {\em Spitzer} IRAC infrared band centered at 3.6\micron. The long-wavelength baseline of our spectrum allows for $K$-corrections to be computed out to $z \sim 3-6$ for the 2MASS filters and $z \sim 10$ for the 3.6 \micron\ band. 

Since our composite spectrum only extends to 3.5\micron\ and the IRAC 3.6\micron\ bandpass extends to $\sim 4$\micron, we must extrapolate the long wavelength end of our spectrum with the power-law fit for that region (derived in \S 4.1) to measure the rest-frame emission through the bandpass and relate redshifted emission to it.  As there are no prominent emission lines expected in the 3.5-4 \micron\ band, this continuum extrapolation should be adequate.  

We define the $K$-correction, $K(z)$, as the difference between the rest-frame absolute magnitude, $M$, of a source and its observed magnitude, $m$, at redshift $z$, through the same bandpass.  Symbolically, 
\begin{equation}
m = M + 5\log_{10}(\frac{\mathrm{D}_\mathrm{L}}{10\mathrm{pc}}) + K(z),
\end{equation}
where D$_\mathrm{L}$ is the luminosity distance.  The $K$-correction is then computed from the source flux per unit wavelength, $f(\lambda)$, and the response function for the bandpass, $T(\lambda)$ with the following equation:

\begin{equation}
K(z) = 2.5\log_{10}(1+z) + 2.5\log_{10}\left[ \frac{\int T(\lambda) f(\lambda) d\lambda}{\int T(\lambda) f(\lambda/(1+z)) d\lambda} \right].
\end{equation}

\begin{figure}
\epsscale{1}
\plotone{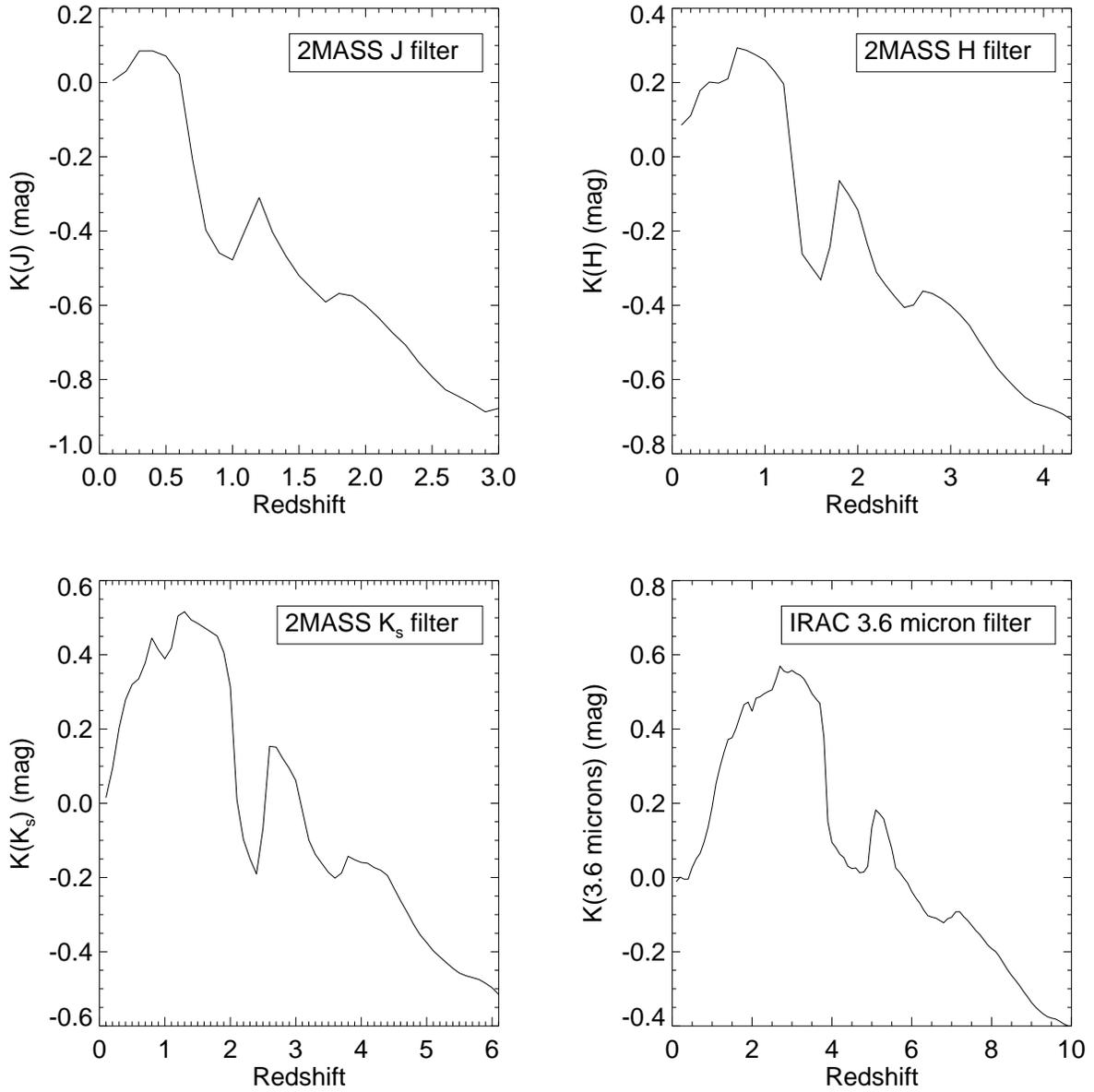}
\caption{$K$-corrections as a function of redshift for the three 2MASS bands and the {\em Spitzer} 3.6\micron\ IRAC band. \label{kcorr}}
\end{figure}

We use the response functions for the 2MASS bands and the {\em Spitzer} 3.6\micron\ IRAC band in equation (7) and compute the $K$-corrections for our composite quasar from $z=0-10$ in increments of $\Delta z = 0.1$.  The results are shown in Figure \ref{kcorr} and Table \ref{kcorrtbl}.

\section{Summary and Future Work}

We have presented the first-ever near-infrared quasar composite spectrum spanning the wavelength range 0.57-3.5\micron.  Combined with optical spectra, this composite extends to 0.3\micron.  We have analyzed the continuum in this wavelength range to reveal three distinct power-laws with breaks around 1.08\micron\ and 2.22\micron.  The optical part of the spectrum (blueward of 1.08 \micron) is best fit by a spectral slope of $\alpha_\nu =-0.78$ while the middle of the spectrum is best fit by $\alpha_\nu =-1.81$.  For wavelengths longer than 2.22 \micron, the spectrum is best fit by $\alpha_\nu =-1.03$.  We find that the order in which we combine our spectra to produce the composite has the strongest effect on the optical spectral index, with the spectrum sorted from high  to low redshift having a spectral index as hard as $\alpha_\nu =-0.37$.  The other spectral regions vary by $\Delta\alpha_\nu \lesssim 0.2$.  An alternative fit combining a single power-law and a blackbody provides an excellent description of the spectrum and offers strong evidence for the existence of hot dust ($T_d \sim 1250$ K) in these blue quasars.

We have also studied the emission lines in this new spectral region, revealing the Paschen series lines as well as oxygen and helium and forbidden sulfur emission.  Paschen line ratios rule out case B recombination in the broad-line region.  The small number of spectra (13) obtained in the LXD band resulted in a modest signal-to-noise ratio beyond 2.1\micron\ for our composite.  Additional spectra in this band will allow measurements of the intensities and equivalent widths of the Brackett series lines expected in this wavelength region.

We have compared radio-quiet and radio-loud subsample composites to the radio-quiet and radio-loud quasar SEDs produced by \citet{Elvis94} .  We find general agreement in the spectral shapes beyond $\sim 1$ \micron.

We have computed $K$-corrections to the 2MASS $J$, $H$, and $K_s$ bands as well as the {\em Spitzer} 3.6\micron\ IRAC band for the purpose of predicting fluxes for high-redshift quasars observed in these bands.  Since IRTF has an LXD mode which captures the 2.3-5.5\micron\ wavelength range, additional spectra in this band could extend our composite (for sources with the same redshift range) out to $\sim 4.5\micron$.  This will allow for a better measurement of the $K$-correction to the 3.6\micron\ IRAC band as well as for the determination of the $K$-correction for the 4.5\micron\ IRAC band when we extrapolate out to 5\micron\ (the extent of the passband).  Such a composite can be combined with future {\em Spitzer} IRS spectra of low-redshift luminous blue quasars planned by guaranteed time observers to extend the wavelength coverage, and our understanding of the intrinsic quasar spectrum, out to $\sim 40\micron$. \\

We thank Martin Elvis for his helpful comments on our comparison with his quasar SED. This work was supported at Columbia by the NSF under grant AST-00-98259.

E. G. is a Visiting Astronomer at the Infrared Telescope Facility, which is operated by the University of Hawaii under Cooperative Agreement no. NCC 5-538 with the National Aeronautics and Space Administration, Office of Space Science, Planetary Astronomy Program.

Funding for the creation and distribution of the SDSS Archive has been provided by the Alfred P. Sloan Foundation, the Participating Institutions, the National Aeronautics and Space Administration, the National Science Foundation, the U.S. Department of Energy, the Japanese Monbukagakusho, and the Max Planck Society. The SDSS Web site is \url{http://www.sdss.org/}.

The SDSS is managed by the Astrophysical Research Consortium (ARC) for the Participating Institutions. The Participating Institutions are The University of Chicago, Fermilab, the Institute for Advanced Study, the Japan Participation Group, The Johns Hopkins University, the Korean Scientist Group, Los Alamos National Laboratory, the Max-Planck-Institute for Astronomy (MPIA), the Max-Planck-Institute for Astrophysics (MPA), New Mexico State University, University of Pittsburg, University of Portsmouth, Princeton University, the United States Naval Observatory, and the University of Washington.

This publication makes use of data products from the Two Micron All Sky Survey, which is a joint project of the University of Massachusetts and the Infrared Processing and Analysis Center/California Institute of Technology, funded by the National Aeronautics and Space Administration and the National Science Foundation.

\appendix
\section{Appendix}

Because near-infrared spectroscopy for the objects used to form the composite are not published in the literature, we make available the twenty-seven fully reduced quasar spectra to the astronomical community.  Table \ref{qspec} displays a sample spectrum of SDSSJ000943.1$-$090839.2.  This spectrum, as well as the other 26 spectra used to create the composite, are available in their entirety in the electronic edition of the Journal.

\pagebreak

\bibliography{/home/yoyoma/eilatg/Pubs/bibtex/eilatg_refs}

\input{tab1}
\input{tab2}
\input{tab3}
\input{tab4}
\input{tab5}
\input{tab6}
\input{tab7}
\input{tab8}
\input{tabA1}

\clearpage

\end{document}

%% file: tab1.tex
\begin{deluxetable}{ccclcclcc}
\tabletypesize{\footnotesize}
\tablewidth{0pt}
\tablehead{
\colhead{R.A.} & \colhead{Dec} & \colhead{$K_s$} &
\multicolumn{3}{c}{SXD} & \multicolumn{3}{c}{LXD} \\
\colhead{(J2000.0)} & \colhead{(J2000.0)} & \colhead{(mag)} &
\colhead{Date} & \colhead{Exp (s)} & \colhead{Slit (\arcsec)} & 
\colhead{Date} & \colhead{Exp (s)} & \colhead{Slit (\arcsec)} \\
\colhead{(1)} & \colhead{(2)} & \colhead{(3)} & 
\colhead{(4)} & \colhead{(5)} & \colhead{(6)} & 
\colhead{(7)} & \colhead{(8)} & \colhead{(9)}
}
\tablecaption{Observation Log\label{observations}}
\startdata
00 09 43.14 & $-$09 08 39.2 & 14.03 & 2004 Sep 13 & 2160  & 0.5   & \nodata     &\nodata &\nodata \\
00 13 27.31 & $+$00 52 31.9 & 14.34 & 2004 Sep 14 & 2160  & 0.5   & \nodata     &\nodata &\nodata \\
00 57 09.92 & $+$14 46 10.1 & 12.12 & 2004 Sep 13 & 1080  & 0.5   & 2004 Sep 14 &  900 & 0.5 \\
00 58 12.85 & $+$16 02 01.3 & 13.15 & 2004 Sep 13 & 1440  & 0.5   & 2004 Sep 14 & 1500 & 0.5 \\
01 02 26.31 & $-$00 39 04.6 & 13.41 & 2003 Sep 25 & 1440  & 0.5   & 2004 Sep 14 & 1500 & 0.5 \\
01 11 10.04 & $-$10 16 31.8 & 13.05 & 2003 Sep 24 & 1440  & 0.5   & 2004 Sep 13 & 1800 & 0.5 \\
01 34 18.19 & $+$00 15 36.6 & 13.55 & 2004 Sep 14 & 1440  & 0.5   & \nodata     &\nodata &\nodata \\
01 55 30.02 & $-$08 57 04.0 & 12.65 & 2003 Sep 23 &  960  & 0.5   & 2004 Sep 14 & 1200 & 0.5 \\
01 59 10.05 & $+$01 05 14.5 & 13.51 & 2003 Sep 23 & 1440  & 0.5   & 2004 Sep 14 & 1800 & 0.5 \\
01 59 50.24 & $+$00 23 40.8 & 12.01 & 2003 Sep 23 &  960  & 0.5   & 2004 Sep 13 & 1200 & 0.5 \\
02 17 07.87 & $-$08 47 43.4 & 14.15 & 2003 Sep 24 & 1440  & 0.5   & \nodata     &\nodata &\nodata \\
02 42 50.85 & $-$07 59 14.2 & 14.16 & 2003 Sep 23 & 1920  & 0.5   & \nodata     &\nodata &\nodata \\
03 12 09.20 & $-$08 10 13.8 & 14.17 & 2004 Sep 13 & 1800  & 0.5   & \nodata     &\nodata &\nodata \\
03 22 13.89 & $+$00 55 13.4 & 12.90 & 2004 Sep 12 & 1080  & 0.5   & \nodata     &\nodata &\nodata \\
03 40 00.50 & $-$05 17 47.1 & 13.59 & 2004 Sep 13 & 1200  & 0.5   & \nodata     &\nodata &\nodata \\
07 36 23.12 & $+$39 26 17.8 & 12.27 & 2004 Nov 24\tablenotemark{a} & 900 & 0.6 &\nodata&\nodata&\nodata \\
12 55 19.69 & $+$01 44 12.2 & 13.43 & 2004 Jun 12 &  960  & 0.8   & 2004 Jun 12 & 1800 & 0.8 \\
13 07 56.57 & $+$01 07 09.6 & 13.57 & \nodata     &\nodata&\nodata& 2004 Jun 13 & 1800 & 0.8 \\
15 06 10.50 & $+$02 16 49.9 & 12.79 & 2004 Jun 12 &  960  & 0.8   & 2004 Jun 12 & 1320 & 0.8 \\
16 05 07.92 & $+$48 34 22.0 & 13.38 & \nodata     &\nodata&\nodata& 2004 Jun 13 & 1800 & 0.8 \\
17 04 41.37 & $+$60 44 30.5 & 12.43 & 2004 Sep 12 &  960  & 0.5   & 2004 Jun 13 &  900 & 0.8 \\
17 27 11.81 & $+$63 22 41.8 & 14.17 & 2004 Sep 14 & 2160  & 0.5   & \nodata     &\nodata &\nodata \\
21 18 43.24 & $-$06 36 18.0 & 14.45 & 2003 Sep 23 & 1920  & 0.5   & 2004 Sep 12 & 2400 & 0.5 \\
21 26 19.65\tablenotemark{b} & $-$06 54 08.9\tablenotemark{b} & 14.39 & 2004 Jun 12 & 1920 &       & \nodata & \nodata \\
            &               &       & 2004 Sep 13 & 1920  & 0.5   & \nodata     &\nodata &\nodata \\
23 22 35.66 & $-$09 44 38.1 & 14.19 & 2004 Sep 14 & 2160  & 0.5   & \nodata     &\nodata &\nodata \\
23 49 32.77 & $-$00 36 45.8 & 14.02 & 2003 Sep 25 & 1440  & 0.5   & 2004 Jun 13 & 1800 & 0.8 \\
23 51 56.12\tablenotemark{b} & $-$01 09 13.3\tablenotemark{b} & 12.18 & 2003 Sep 23 & 1440 & 0.5 & 2004 Jun 12 & 1080 & 0.8\\
            &               &        & \nodata    &\nodata&\nodata& 2004 Sep 12 & 1800 & 0.3 \\
\enddata
\tablenotetext{a}{Spectrum from the TIFKAM camera at the MDM observatory.}
\tablenotetext{b}{Data from observations taken at two epochs were combined into one spectrum.}
\end{deluxetable}

%% file: tab2.tex
\begin{deluxetable}{cccccccccccccc}
\tabletypesize{\footnotesize}
\rotate
\tablewidth{0pt}
\tablehead{
\colhead{R.A.} & \colhead{Dec} & \colhead{$z$\tablenotemark{a}} & 
\colhead{$u$} & \colhead{$g$} & \colhead{$r$} & \colhead{$i$} & \colhead{$z$} &
\colhead{$J$} & \colhead{$H$} & \colhead{$K_s$} & 
\colhead{$F_{pk}$} & \colhead{$R_r$\tablenotemark{d}} & \colhead{$M_i$} \\
\colhead{(J2000.0)} & \colhead{(J2000.0)} & \colhead{} & 
\colhead{(mag)} & \colhead{(mag)} & \colhead{(mag)} & \colhead{(mag)} & \colhead{(mag)} & 
\colhead{(mag)\tablenotemark{b}} & \colhead{(mag)\tablenotemark{b}} & \colhead{(mag)\tablenotemark{b}} & 
\colhead{(mJy)} & \colhead{} & \colhead{(mag)}\\
\colhead{(1)} & \colhead{(2)} & \colhead{(3)} & 
\colhead{(4)} & \colhead{(5)} & \colhead{(6)} & \colhead{(7)} & \colhead{(8)} &
\colhead{(9)} & \colhead{(10)} & \colhead{(11)} & 
\colhead{(12)} & \colhead{(13)} & \colhead{(14)}
}
\tablecaption{Summary of Quasar Properties\label{targets}}
\startdata
00 09 43.14 & $-$09 08 39.2 & 0.210 & 17.46 & 17.40 & 17.29 & 16.92 & 17.03 & 15.67 & 14.92 & 14.03 &\nodata&\nodata& $-$23.12\\
00 13 27.31 & $+$00 52 31.9 & 0.362 & 18.07 & 17.73 & 17.56 & 17.60 & 17.05 & 16.21 & 15.50 & 14.34 &\nodata&\nodata& $-$23.72\\
00 57 09.92 & $+$14 46 10.1 & 0.172 & 16.24 & 16.11 & 15.99 & 15.42 & 15.63 & 14.11 & 13.23 & 12.12 & 2.4\tablenotemark{c}& 1.6 & $-$24.19\\
00 58 12.85 & $+$16 02 01.3 & 0.211 & 17.82 & 17.13 & 16.63 & 16.23 & 16.28 & 15.05 & 14.12 & 13.15 &\nodata&\nodata& $-$23.91\\
01 02 26.31 & $-$00 39 04.6 & 0.295 & 16.14 & 16.10 & 15.99 & 16.15 & 15.75 & 15.13 & 14.39 & 13.41 &\nodata&\nodata& $-$24.70\\
01 11 10.04 & $-$10 16 31.8 & 0.179 & 17.13 & 17.09 & 16.76 & 16.38 & 16.45 & 14.78 & 13.94 & 13.05 &   9.33&  13.0 & $-$23.29\\
01 34 18.19 & $+$00 15 36.6 & 0.400 & 16.97 & 16.81 & 16.81 & 16.77 & 16.40 & 15.61 & 14.77 & 13.55 &\nodata&\nodata& $-$24.79\\
01 55 30.02 & $-$08 57 04.0 & 0.165 & 17.00 & 16.97 & 16.80 & 16.31 & 16.45 & 14.71 & 13.68 & 12.65 &\nodata&\nodata& $-$23.15\\
01 59 10.05 & $+$01 05 14.5 & 0.217 & 17.48 & 17.44 & 17.24 & 16.95 & 17.05 & 15.42 & 14.53 & 13.51 &\nodata&\nodata& $-$23.16\\
01 59 50.24 & $+$00 23 40.8 & 0.163 & 15.91 & 15.88 & 15.97 & 15.75 & 15.82 & 14.07 & 13.09 & 12.01 &  22.55&  15.2 & $-$23.69\\
02 17 07.87 & $-$08 47 43.4 & 0.292 & 17.78 & 17.41 & 17.14 & 17.13 & 16.85 & 15.77 & 15.08 & 14.15 &\nodata&\nodata& $-$23.67\\
02 42 50.85 & $-$07 59 14.2 & 0.378 & 17.23 & 17.06 & 17.03 & 17.09 & 16.71 & 15.88 & 15.12 & 14.16 &\nodata&\nodata& $-$24.33\\
03 12 09.20 & $-$08 10 13.8 & 0.265 & 17.72 & 17.51 & 17.32 & 17.24 & 16.87 & 15.98 & 15.16 & 14.17 &\nodata&\nodata& $-$23.40\\
03 22 13.89 & $+$00 55 13.4 & 0.185 & 17.08 & 16.86 & 16.61 & 16.14 & 16.34 & 14.81 & 14.01 & 12.90 &\nodata&\nodata& $-$23.78\\
03 40 00.50 & $-$05 17 47.1 & 0.217 & 18.12 & 18.00 & 17.76 & 17.76 & 17.35 & 16.23 & 15.39 & 14.22 &\nodata&\nodata& $-$23.49\\
07 36 23.12 & $+$39 26 17.8 & 0.118 & 16.26 & 16.39 & 16.25 & 15.49 & 15.77 & 14.18 & 13.33 & 12.27 &   3.50&   3.0 & $-$23.26\\
12 55 19.69 & $+$01 44 12.2 & 0.343 & 16.02 & 16.01 & 16.05 & 16.25 & 15.52 & 15.23 & 14.47 & 13.43 &   1.35&   1.0 & $-$24.93\\
13 07 56.57 & $+$01 07 09.6 & 0.276 & 18.10 & 17.58 & 17.15 & 17.06 & 16.50 & 15.44 & 14.67 & 13.57 &\nodata&\nodata& $-$23.61\\
15 06 10.50 & $+$02 16 49.9 & 0.135 & 16.47 & 16.39 & 16.32 & 15.85 & 15.98 & 14.68 & 13.72 & 12.79 &\nodata&\nodata& $-$23.21\\
16 05 07.92 & $+$48 34 22.0 & 0.295 & 16.66 & 16.69 & 16.68 & 16.80 & 16.28 & 15.27 & 14.41 & 13.38 &\nodata&\nodata& $-$24.01\\
17 04 41.37 & $+$60 44 30.5 & 0.372 & 15.65 & 15.44 & 15.25 & 15.18 & 14.84 & 14.15 & 13.47 & 12.43 &\nodata&\nodata& $-$26.19\\
17 27 11.81 & $+$63 22 41.8 & 0.218 & 17.38 & 17.19 & 17.02 & 16.74 & 16.74 & 15.54 & 14.92 & 14.17 &   1.36&   2.4 & $-$23.37\\
21 18 43.24 & $-$06 36 18.0 & 0.328 & 17.65 & 17.67 & 17.59 & 17.69 & 16.80 & 16.38 & 15.65 & 14.45 &  75.52& 225.1 & $-$23.66\\
21 26 19.65 & $-$06 54 08.9 & 0.418 & 17.16 & 16.92 & 16.85 & 16.76 & 16.62 & 15.94 & 15.27 & 14.39 &  69.80& 105.6 & $-$25.04\\
23 22 35.66 & $-$09 44 38.1 & 0.372 & 17.11 & 16.99 & 17.05 & 17.05 & 16.55 & 16.12 & 15.23 & 14.19 &\nodata&\nodata& $-$24.33\\
23 49 32.77 & $-$00 36 45.8 & 0.279 & 17.38 & 17.25 & 17.12 & 17.20 & 16.84 & 15.97 & 15.28 & 14.02 &\nodata&\nodata& $-$23.50\\
23 51 56.12 & $-$01 09 13.3 & 0.174 & 15.68 & 15.76 & 15.79 & 15.37 & 15.59 & 14.33 & 13.41 & 12.18 & 166.50&  94.6 & $-$24.23\\
\enddata
\tablenotetext{a}{Redshifts are from the SDSS quasar catalog \citep{Schneider03}.}
\tablenotetext{b}{2MASS measurements are Vega-based, not AB, magnitudes.}
\tablenotetext{c}{Flux density from NVSS \citep{Condon98}.}
\tablenotetext{d}{Radio loudness parameter, defined by $R_r = f_\mathrm{radio}/f_r$, where $f_r = 3631\times10^{-0.4r}$ Jy. }
\end{deluxetable}

%% file: tab3.tex
\begin{deluxetable}{rccc}
\tabletypesize{\footnotesize}
\tablewidth{0pt}
\tablehead{
\colhead{$\lambda$} & \colhead{$<f_{\lambda}>$}  &
\colhead{$<f_{\lambda}>$ Uncertainty} & 
\colhead{$<f_{\lambda}>_\mathrm{gm}$}  \\
\colhead{(\AA)} & \colhead{(arbitrary units)}  &
\colhead{(arbitrary units)} & 
\colhead{(arbitrary units)} 
}
\tablecaption{Composite Quasar Spectrum, Arithmetic and Geometric Means\label{spectbl}}
\startdata
 5900.4 & 1.9824 & 0.1895 & 2.2860\\
 5901.7 & 2.0507 & 0.2737 & 2.4589\\
 5903.1 & 2.6955 & 0.2954 & 3.0849\\
 5904.5 & 1.8391 & 0.2292 & 2.5981\\
 5905.9 & 3.2544 & 0.2552 & 3.0030\\
 5907.2 & 2.6429 & 0.2391 & 2.6205\\
 5908.6 & 2.4293 & 0.1401 & 2.4708\\
 5910.0 & 2.6892 & 0.1674 & 1.4671\\
 5911.4 & 2.5463 & 0.1675 & 2.6436\\
 5912.7 & 2.1509 & 0.1073 & 1.1448\\
\enddata
\tablecomments{The complete version of this table is in the electronic edition of the Journal.  The printed edition contains only a sample.}
\end{deluxetable}

%% file: tab4.tex
\begin{deluxetable}{cccccccccc}
\tabletypesize{\footnotesize}
\rotate
\tablewidth{0pt}
\tablehead{
\colhead{Spectrum} & \colhead{UV - Optical} & \colhead{$\alpha_\nu$} &
\colhead{H-$\alpha$} & \colhead{$\alpha_\nu$} &
\colhead{mid-infrared} & \colhead{$\alpha_\nu$} &
\colhead{far-infrared} & \colhead{$\alpha_\nu$} & \colhead{$\chi^2$} \\
\colhead{(1)} & \colhead{(2)} & \colhead{(3)} & 
\colhead{(4)} & \colhead{(5)} & \colhead{(6)} & 
\colhead{(7)} & \colhead{(8)} & \colhead{(9)} & \colhead{(10)}
}
\tablecaption{Fitted Power-Law Indices\label{spec_index}}
\startdata
1 &Ly$\alpha$ - H$\beta$&$-0.46$& \nodata          & \nodata       & \nodata &\nodata & \nodata &\nodata & \nodata \\
2 &1350-4230\AA         &$-0.44$&$6005-7180\phn$\AA& $-2.45$       & \nodata &\nodata & \nodata &\nodata & \nodata \\
3 &\nodata              &\nodata&$6005-7180\phn$\AA& $-1.03\pm0.01$& \nodata &\nodata & \nodata &\nodata & \nodata \\
4 &\nodata              &\nodata&$6005-7180\phn$\AA& $-1.55\pm0.01$& \nodata &\nodata & \nodata &\nodata & \nodata \\
5 &\nodata              &\nodata&$6005-7180\phn$\AA& $-1.21\pm0.01$& \nodata &\nodata & \nodata &\nodata & \nodata \\
6 &\nodata              &\nodata&$5700-10850$\AA   & $-0.78$       & 1.085-2.230 \micron & $-1.81$ & 2.23-3.50 \micron & $-1.03$ &  3.39 \\
7 &\nodata              &\nodata&$5700-9570\phn$\AA& $-0.37$       & 0.957-2.093 \micron & $-1.73$ & 2.09-3.50 \micron & $-0.96$ &  3.39 \\
8 &\nodata              &\nodata&$5700-9730\phn$\AA& $-0.48$       & 0.973-2.382 \micron & $-1.74$ & 2.38-3.50 \micron & $-1.17$ &  3.36 \\
\hline
%9 &\nodata              &\nodata&$5700-8960\phn$\AA& $-0.16$       & 0.896-1.808 \micron & $-1.65$ & 1.81-3.50 \micron & $-1.48$ & 11.51\\
%10&\nodata              &\nodata&$5700-12440$\AA   & $-1.19$       & 1.244-1.837 \micron & $-1.96$ & 1.84-3.50 \micron & $-1.16$ & 10.03\\
%\hline
%11&\nodata              &\nodata&$5700-12610$\AA   & $-0.39$       & 1.261-2.264 \micron & $-2.18$ & 2.26-3.50 \micron & $-1.17$ & \nodata \\
%12&\nodata              &\nodata&$5700-12180$\AA   & $-0.15$       & 1.218-1.860 \micron & $-1.64$ & 1.86-3.50 \micron & $-1.20$ & \nodata \\
\enddata
\tablerefs{(1) \citet{Brotherton01}; (2) \citet{VandenBerk01}; (3) This paper, optical composite ordered by $z_{low} - z_{high}$; (4) This paper, optical composite ordered by $z_{high} - z_{low}$; (5) This paper, optical composite ordered by $z_{high~snr} - z_{low~snr}$; (6) This paper, near-infrared composite ordered by $z_{low} - z_{high}$; (7) This paper, near-infrared composite ordered by $z_{high} - z_{low}$; (8) This paper, near-infrared composite ordered by $z_{high~snr} - z_{low~snr}$}%; (9) This paper, radio-quiet subsample; (10) This paper, radio-loud subsample; (11) \citet{Elvis94} radio-quiet SED; (12) \citet{Elvis94} radio-loud SED.}
\end{deluxetable}

%% file: tab5.tex
\begin{deluxetable}{cccc}
\tabletypesize{\footnotesize}
\tablewidth{0pt}
\tablehead{
\colhead{Spectrum} & \colhead{$\alpha_\nu$} & \colhead{$T_{dust}$ (K)} & \colhead{$\chi^2$}\\

\colhead{(1)} & \colhead{(2)} & \colhead{(3)} & \colhead{(4)} 
}
\tablecaption{Fitted Power-Law plus Blackbody\label{bb}}
\startdata
1               &-0.92 & 1260 & 3.18\\
2               &-0.94 & 1260 & 3.21\\
3               &-0.93 & 1250 & 3.17\\
\hline
4               & -1.34 & 1260 & 3.55               \\
                & -0.92\tablenotemark{a} & 1230 & 2.96 \\
                & -0.80 & 1260\tablenotemark{a} & 2.93 \\
5               & -0.92\tablenotemark{a} & 1330 & 4.30 \\
                & -0.85 & 1260\tablenotemark{a} & 4.64 \\
\enddata
\tablenotetext{a}{This parameter is held fixed in the fit.}
\tablerefs{(1) Near-infrared composite ordered by $z_{low} - z_{high}$; (2) Near-infrared composite ordered by $z_{high} - z_{low}$; (3) Near-infrared composite ordered by $z_{high~snr} - z_{low~snr}$; (4) Radio-quiet subsample; (5) Radio-loud subsample.}
\end{deluxetable}

%% file: tab6.tex
\begin{deluxetable}{cccrrcrr}
\tabletypesize{\footnotesize}
\tablewidth{0pt}
\tablehead{
\colhead{Emission Feature} & \colhead{Lab $\lambda$} & \colhead{Measured $\lambda$} &
\colhead{Start\tablenotemark{a}} & \colhead{End\tablenotemark{a}} &
\colhead{Intensity} & \colhead{EW} & \colhead{FWHM} \\
\colhead{} & \colhead{(\AA)} & \colhead{(\AA)} &
\colhead{(\AA)} & \colhead{(\AA)} & 
\colhead{$[100 \times F/F(\mathrm{H}_\alpha)]$} & \colhead{(\AA)} & \colhead{(\AA)} \\
\colhead{(1)} & \colhead{(2)} & \colhead{(3)} & 
\colhead{(4)} & \colhead{(5)} & 
\colhead{(6)} & \colhead{(7)} & \colhead{(8)} 
}
\tablecaption{Emission Line Features\label{linetbl}}
\startdata
H$\alpha$                    &\phn6563 &\phn6569.5$\pm$0.3 & 6463.6&  6674.5&      100.00$\pm$0.98& 180.4 & 47.4\\
\ion{O}{1}                   &\phn8446 &\phn8457.5$\pm$0.3 & 8228.7&  8684.9&\phn\phn3.78$\pm$0.16&  10.7 &104.2\\
\ [\ion{S}{3}]               &\phn9069 &\phn9076.8$\pm$0.8 & 9054.1&  9098.4&\phn\phn0.25$\pm$0.04&   0.8 & 19.0\\
\ion{Fe}{2}                  &\phn9202 &\phn9214.0$\pm$1.4 & 9115.4&  9312.5&\phn\phn1.10$\pm$0.09&   3.5 & 81.4\\
Pa$\epsilon$                 &\phn9545 &\phn9534.4$\pm$0.5 & 9443.4&  9627.5&\phn\phn2.02$\pm$0.09&   7.0 & 39.9\\
Pa$\delta$                   &   10049 & 10042.3$\pm$0.3 & 9688.1& 10395.5&\phn\phn5.72$\pm$0.13&  21.1 &161.2\\
\ion{He}{1}\tablenotemark{b} &   10830 & 10830.0\phs\phn\phd\phn&10635.2& 11025.6&   \phn10.05$\pm$0.20&  36.0 &115.8\\
Pa$\gamma$\tablenotemark{b}  &   10941 & 10941.0\phs\phn\phd\phn&10702.2&11178.0& \phn\phn1.95$\pm$0.71&   7.0 &109.4\\
\ion{O}{1}                   &   11287 & 11296.4$\pm$1.3 &11211.8& 11380.0&\phn\phn0.89$\pm$0.04&   3.3 & 78.8\\
Pa$\beta$                    &   12820 & 12821.3$\pm$0.1 &12533.2& 13108.7&\phn\phn4.82$\pm$0.06&  18.4 &128.3\\
Pa$\alpha$                   &   18756 & 18735.5$\pm$2.1 &18499.5& 18970.1&\phn\phn3.10$\pm$0.01&  12.7 &196.1\\
\ion{He}{1}                  &   20580 & 20506.8$\pm$6.4 &20035.8& 20975.2&\phn\phn1.89$\pm$0.03&   8.6 &300.5\\
\enddata
\tablenotetext{a}{Wavelengths between which the line is integrated.  The limits are chosen at $5\%$ of the peak, above the continuum.}
\tablenotetext{b}{The line centers were fixed when fitting \ion{He}{1} and Pa$\gamma$ due to the strong blending of these lines.}
\end{deluxetable}

%% file: tab7.tex
\begin{deluxetable}{rccc}
\tabletypesize{\footnotesize}
\tablewidth{0pt}
\tablehead{
\colhead{$\lambda$} & \colhead{$<f_{\lambda}>$}  &
\colhead{$<f_{\lambda}>$ Uncertainty} & 
\colhead{$<f_{\lambda}>_\mathrm{gm}$}  \\
\colhead{(\AA)} & \colhead{(arbitrary units)}  &
\colhead{(arbitrary units)} & 
\colhead{(arbitrary units)} 
}
\tablecaption{Optical-to-near-infrared Composite Quasar Spectrum, Arithmetic and Geometric Means\label{optirspectbl}}
\startdata
 2761.6 & 12.2489 & 0.3339 & 8.9927\\
 2762.2 & 11.9368 & 0.3186 &10.7421\\
 2762.9 & 12.0142 & 0.3266 &10.7801\\
 2763.5 & 12.5634 & 0.3259 &11.1975\\
 2764.2 & 12.4229 & 0.3171 &11.0970\\
 2764.8 & 11.8648 & 0.3089 &10.7094\\
 2765.5 & 11.9007 & 0.3047 &10.6902\\
 2766.1 & 11.9286 & 0.3036 &10.6816\\
 2766.8 & 11.9703 & 0.3048 &10.7700\\
 2767.4 & 12.3445 & 0.3069 &11.0721\\
\enddata
\tablecomments{The complete version of this table is in the electronic edition of the Journal.  The printed edition contains only a sample.}
\end{deluxetable}

%% file: tab8.tex
\begin{deluxetable}{crrrr}
\tabletypesize{\footnotesize}
\tablewidth{0pt}
\tablehead{
\colhead{Redshift} & \colhead{$K(J)$}  &
\colhead{$K(H)$} & \colhead{$K(K_s)$} &
\colhead{$K(3.6 \micron)$} \\
}
\tablecaption{Quasar $K$-corrections\label{kcorrtbl}}
\startdata
 0.1 &   0.005 &   0.085 &   0.015 &$-$0.012\\
 0.2 &   0.030 &   0.111 &   0.094 &   0.001\\
 0.3 &   0.085 &   0.178 &   0.202 &$-$0.005\\
 0.4 &   0.085 &   0.201 &   0.279 &$-$0.004\\
 0.5 &   0.071 &   0.199 &   0.320 &   0.027\\
 0.6 &   0.022 &   0.210 &   0.335 &   0.050\\
 0.7 &$-$0.208 &   0.294 &   0.378 &   0.065\\
 0.8 &$-$0.397 &   0.287 &   0.446 &   0.095\\
 0.9 &$-$0.459 &   0.275 &   0.414 &   0.136\\
 1.0 &$-$0.478 &   0.260 &   0.389 &   0.190\\
 1.1 &$-$0.393 &   0.231 &   0.418 &   0.253\\
 1.2 &$-$0.310 &   0.196 &   0.505 &   0.299\\
 1.3 &$-$0.403 &$-$0.033 &   0.516 &   0.338\\
 1.4 &$-$0.466 &$-$0.262 &   0.494 &   0.372\\
 1.5 &$-$0.520 &$-$0.297 &   0.485 &   0.377\\
\enddata
\tablecomments{The complete version of this table is in the electronic edition of the Journal.  The printed edition contains only a sample.}
\end{deluxetable}

%% file: tabA1.tex
%% The values (usually only l,r and c) in the last part of
%% \begin{deluxetable}{} command tell LaTeX how many columns
%% there are and how to align them.
\begin{deluxetable}{ccc}
\tablewidth{0pt}

%% Keep a portrait orientation

%% Over-ride the default font size
%% Use Default (12pt)

%% Use \tablewidth{?pt} to over-ride the default table width.
%% If you are unhappy with the default look at the end of the
%% *.log file to see what the default was set at before adjusting
%% this value.

%% This is the title of the table.
\tablecaption{Spectrum of SDSSJ000943.1$-$090839.2\label{qspec}}

%% This command over-rides LaTeX's natural table count
%% and replaces it with this number.  LaTeX will increment 
%% all other tables after this table based on this number
\tablenum{A1}

%% The \tablehead gives provides the column headers.  It
%% is currently set up so that the column labels are on the
%% top line and the units surrounded by ()s are in the 
%% bottom line.  You may add more header information by writing
%% another line between these lines. For each column that requries
%% extra information be sure to include a \colhead{text} command
%% and remember to end any extra lines with \\ and include the 
%% correct number of &s.
\tablehead{\colhead{$\lambda$} & \colhead{$f_\lambda$} & \colhead{$f_\lambda$ Uncertainty} \\ 
\colhead{(\AA)} & \colhead{(arbitrary units)} & \colhead{(arbitrary units)} } 

%% All data must appear between the \startdata and \enddata commands
\startdata
0.807880 & $-$1.07676e$-$16 & 8.51939e$-$17 \\
0.808083 & $-$3.07219e$-$16 & 1.19947e$-$16 \\
0.808285 & $-$3.00444e$-$16 & 4.82051e$-$16 \\
0.808488 &\phs4.34456e$-$16 & 2.07032e$-$16 \\
0.808690 &\phs4.09549e$-$16 & 3.45445e$-$16 \\
0.808893 &\phs2.37477e$-$16 & 2.07997e$-$16 \\
0.809095 &\phs3.07857e$-$17 & 1.89639e$-$16 \\
0.809298 &\phs2.68657e$-$17 & 3.91788e$-$16 \\
0.809500 & $-$2.09553e$-$17 & 1.89665e$-$16 \\
0.809702 & $-$6.93002e$-$17 & 9.67822e$-$17 \\
0.809905 & $-$1.22391e$-$16 & 3.68280e$-$16 \\
0.810107 &\phs1.84263e$-$16 & 4.10956e$-$16 \\
0.810310 &\phs6.95212e$-$16 & 1.07431e$-$15 \\
0.810512 & $-$5.50372e$-$17 & 4.67514e$-$16 \\
0.810715 & $-$3.68670e$-$17 & 5.78469e$-$17 \\
\enddata

%% Include any \tablenotetext{key}{text}, \tablerefs{ref list},
%% or \tablecomments{text} between the \enddata and 
%% \end{deluxetable} commands

%% General table comment marker
\tablecomments{This spectrum, as well as the other 26 spectra used to create the composite, are available in their entirety in the electronic edition of the Journal.  The printed edition contains only this sample.}

\end{deluxetable}